\documentclass[a4paper,twocolumn,epjc3]{svjour3}          

\RequirePackage[T1]{fontenc}

\smartqed  

\RequirePackage{graphicx}
\RequirePackage{mathptmx}      
\RequirePackage{flushend}
\RequirePackage[numbers,sort&compress]{natbib}
\RequirePackage[colorlinks,citecolor=blue,urlcolor=blue,linkcolor=blue]{hyperref}

\journalname{Eur. Phys. J. C}

\RequirePackage{orcidlink}
\RequirePackage{multirow}
\graphicspath{{fig/}}

\begin{document}

\title{Positron Channeling in Quasi-Mosaic Bent Crystals: Atomistic Simulations vs. Experiment}


\author{Maykel M\'arquez-Mijares\thanksref{e1,addr1}\orcidlink{0000-0003-2503-2802}
        \and
        Germ\'an Rojas-Lorenzo\thanksref{e2,addr1}\orcidlink{0000-0002-7605-9426}
        \and
        Paulo E. Iba\~{n}ez-Almaguer\thanksref{e3,addr1}\orcidlink{0009-0004-9878-3900}
        \and
        Jes\'us Rubayo-Soneira\thanksref{e4,addr1,t1}\orcidlink{0000-0001-6093-9181}
        \and
        Andrey V. Solov'yov\thanksref{e5,addr2,t1}\orcidlink{0000-0003-1602-6144} 
}

\thankstext[$\star$]{t1}{Corresponding author}
\thankstext{e1}{e-mail: mmarquez@instec.cu}
\thankstext{e2}{e-mail: grojas@instec.cu}
\thankstext{e3}{e-mail: paulo.ibanez@instec.cu}
\thankstext{e4}{e-mail: jrubayo@gmail.com}
\thankstext{e5}{e-mail: solovyov@mbnresearch.com}

\institute{Instituto Superior de Tecnolog\'ias y Ciencias Aplicadas, University of Havana (InSTEC-UH). Ave. Salvador Allende 1110, Plaza de la Revoluci\'on. Havana - 10400, Cuba\label{addr1}
          \and
          MBN Research Center. Altenh\"oferallee 3. Frankfurt am Main - 60438, Germany\label{addr2}
}

\date{Received: date / Accepted: date}

\maketitle

\begin{abstract}
This paper reports on a comprehensive study of an ultra-relativistic positron beam deflection by  an oriented quasi-mosaic crystal. The analysis was carried out for the positron energy of 530 MeV incident on the quasi-mosaic bent Si(111) crystal. This particular case was chosen because it has recently been studied experimentally at the \linebreak{} Mainz Microtron MAMI.  The results of the relativistic  mo\-lec\-u\-lar dynamics simulations were  compared with the experimental observations and a good agreement was found. The presence of planar channeling, de-channeling, volume reflection and volume capture processes in the angular distribution of deflected positrons for different beam-crystal alignments has been studied. Predictions have been made for certain crystal orientations for which experimental data are lacking.
\end{abstract}


\section{Introduction}\label{intro}
Inside a crystalline solid, the positions of the constituent atoms follow a regular geometric pattern. This gives rise to an ordered and periodic structure of the crystalline environment, which is slightly disordered by the presence of the atoms' thermal vibrations. As a result, the atoms in the crystal are arranged in  crystalline rows and planes. Therefore, when a high-energy charged particle collides with a crystal, the interaction between the particle and the crystal depends significantly on the orientation of the crystal with respect to the direction of the particle's motion. In this context, the seminal work of Lindhard  \cite{Lindhard_1965} showed that the ordered arrangement of atoms within a crystal generates an electrostatic field that significantly affects the motion of charged particles passing through the crystal at small angles with respect to the crystallographic planes or axes. This effect is known as channeling, in which particles that move along a crystallographic plane (or an axis) undergo correlated interactions
with the atoms that make it up.

The study of the transport of ultra-relativistic charged particle beam through oriented crystals, including the phenomenon of channeling, has become a field of broad scope and investigation  \cite{Biryukov_1997,Uggerhoj_2005,Korol_2014,Sytov_2019,Korol_2022}. The extensive theoretical and experimental studies of channeling and other related phenomena have provided many valuable insights \cite{Korol_2021,Haurylavets_2021,Sushko_2022,Bandiera_2013_1,Sytov_2016,Wistisen_2016,Scandale_2020,Ivanov_2006,Bandiera_2021_1}. 

This paper is devoted to the study of the phenomena that occur when an oriented quasi-mosaic bent  Si(111) crystal is exposed to a well collimated positron beam. 
Positrons entering the crystal nearly parallel to the atomic planes can be channeled through the crystal and follow the crystal curvature.
This process is called planar channeling (CH). As the positrons propagate through the crystal in the channeling regime, scattering events can lead to dechanneling (DCH), 
i.e. the process by which the positrons are thrown out of the channel.

If the crystal is tilted at a small angle with respect to the positron beam, the positron trajectories can be aligned with the crystalographic planes within the crystal. This situation can lead to the volume reflection (VR) events, where positrons are reflected from the bent atomic planes, or to the volume capture (VC) events, where positrons are captured by the bent planar channels and continue their motion in the channeling regime. Recent theoretical and experimental contributions explain these processes very well and describe them with sufficient accuracy \cite{Korol_2014,Korol_2021,Haurylavets_2021,Scandale_2019_1,Mazzolari_2024,Paulo_2024}.

In recent years, a number of experiments have been carried out at various accelerator facilities with the primary aim of exploring the specific features of the channeling process for different crystalline samples and the emission of radiation in bent and periodically bent crystals \cite{Guidi_2012,Bandiera_2013_2,Bagli_2014,Nielsen_2019,Mazzolari_2014,Wienands_2015,Scandale_2019_2,Bandiera_2015,Wistisen_2016,Bagli_2017,Sytov_2017,Wienands_2017,Bandiera_2021_2}.

The present study explores the intricate dynamics of \linebreak{} ultra-relativistic positrons as they propagate through a crystalline medium and interact with the crystal's electrostatic field. To unravel the complexity of this phenomenon, the powerful technique of relativistic molecular dynamics  \cite{Sushko_2013} is used, taking advantage of the state-of-the-art  capabilities of the MBN Explorer software package  \cite{Solovyov_2012,Solovyov_2017,MBN_Explorer,Sushko_2019}. A comprehensive understanding of the problem has been achieved by rigorously comparing the results of the simulations with the recently reported experimental results  \cite{Mazzolari_2024}, obtained for a bent silicon crystal  exposed to irradiation with the 530 MeV positron beam recently developed at the Mainz Microtron MAMI.

The relativistic molecular dynamics approach used in this work is one of a set of algorithms implemented in MBN Explorer, the advanced software package for multiscale simulations of complex molecular structure and dynamics. The case study reported in this work is  in line with the recent roadmap paper on multiscale theory, simulation, and experiment in radiation-exposed condensed matter systems \cite{Solovyov_2024}.

The study of the behavior of charged ultra-relativistic particles within crystalline structures is of immense importance, as it underpins a wide range of applications, from particle accelerator technologies to advances in materials science and beyond. As is well known, the interaction between charged particles incident on a crystalline medium allows us to understand the fundamental mechanisms that govern their dynamics. This research aims to unravel the details of this dynamics.

By exploiting  the synergy between the advanced computational modeling and experimental results, this work aims is to push the frontiers of knowledge in this field and pave the way for innovative applications and further scientific discoveries.

Section \ref{Theory} provides an overview of the methodology used in this study, including the target parameters and the details of the simulations. Section \ref{Results} presents the numerical results obtained and, where possible, compares them with the experimental data collected at MAMI. It also analyses the deflection of the positron beam as a function of  the parameters characterizing  the crystal geometry and the divergence of the incident beam. This analysis is based on the atomistic simulations of positron trajectories using the relativistic molecular dynamics approach. Section \ref{Conclusion} summarises the conclusions of this work and outlines future prospects.
\section{Methodology}\label{Theory}

Following previous  research  \cite{Korol_2014,Korol_2022,Sushko_2022,German_2024,Paulo_2024}, a rigorous study is carried out by generating a significant number of statistically independent trajectories for the projectile pos\-i\-trons. This approach captures  a wide range of possible behaviors and outcomes. The method of relativistic molecular dynamics, implemented in the MBN Explorer package  \cite{Sushko_2013}, is used to simulate the motion of ultra-relativistic positrons in the electrostatic field of a crystalline medium.

Once the dynamic simulation is complete, an advanced analysis of the trajectories is performed. This analysis provides a detailed quantitative characterisation of the particle motion, allowing  us to uncover all the details of their behavior within the crystalline environment.

To model the motion of an ultra-relativistic particle with mass ($m$), charge ($q$), and energy ($\varepsilon$) in an atomic environment, numerical integration of the following relativistic \linebreak{}equations of motion is performed:

\begin{eqnarray}
	\dot{\vec{r}} = \vec{v},
	\qquad
	\dot{\vec{v}} =
	\frac{q}{m \gamma}
	\left(
	\vec{E} -
	\frac{\vec{v} (\vec{E} \cdot \vec{v})}{c^2}
	\right).
	\label{eq:01} 
\end{eqnarray}

In these equations, $\vec{r} = \vec{r}(t)$ and $\vec{v} = \vec{v}(t)$ are the instantaneous coordinate  and  the velocity of the projectile positron, respectively. A dot above the letter indicates a differentiation with respect to time $t$.  The momentum $\vec{p}$  with respect to the velocity is  $\vec{p}=m\gamma \vec{v}(t)$ where $\gamma =\varepsilon /mc^2$  is  the relativistic factor, where $c$ is the speed of light in a vacuum. The electric field at point $\vec{r}$ is calculated as $\vec{E} = -\nabla \phi(\vec{r})$, where $\phi(\vec{r})$  is the potential of the field. This potential is obtained as the sum of the potentials of individual atoms located at points $\vec{r}_i$:

\begin{equation}
	\phi(\vec{r}) = \sum_i \phi_{at}(|\vec{r}-\vec{r}_i|)
\end{equation}

With this approach, the simulation and analysis of motion of the ultra-relativistic positrons in the atomic environment is carried out, taking into account the interaction with the electrostatic field generated by the individual atoms. The simulations with MBN Explorer have been performed with either the Molière potential  \cite{Moliere_1948} or the Fernandez-Pacios potential  \cite{Pacios_1993}. Since the results obtained with both potentials were similar, only the Molière approximation is discussed  below.

The positions of the atoms are generated taking into account the random displacement of the nodes due to thermal vibrations at the room temperature. In addition, the transverse coordinates and velocities of the projectile positrons at the entrance to the crystal are randomly generated according to the chosen value of the beam size and its divergence. These algorithms ensure that each positron trajectory passes  through a unique crystalline environment. It is therefore statistically independent of any other simulated trajectory. A large set of  simulated trajectories correctly reproduces the initial conditions that experienced by the beam positrons as they enter the crystal.

\begin{figure}
	
	\centering 
	\includegraphics[width = 7.cm]{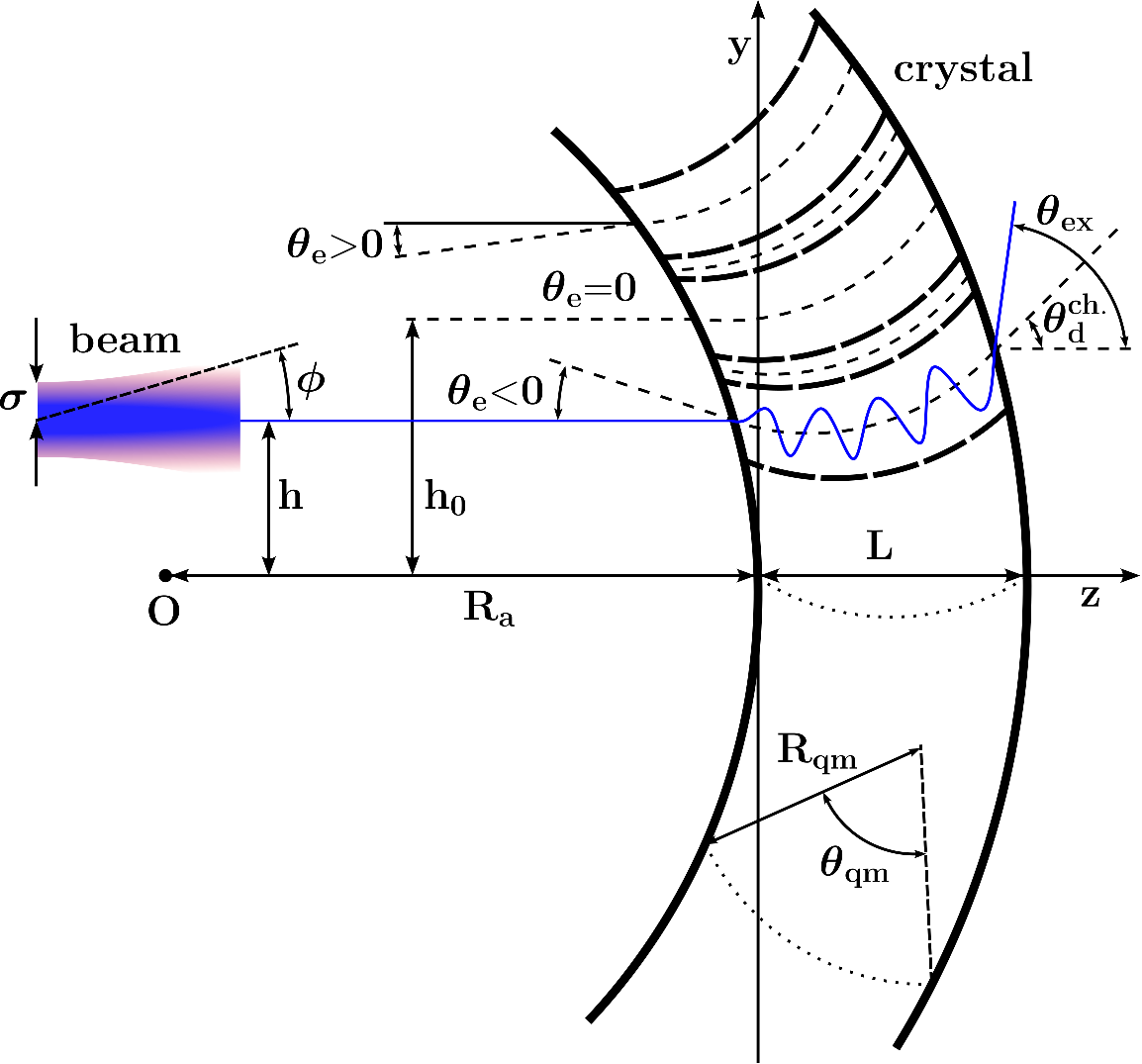}\\
	\caption{Schematic representation of a `quasi-mosaic' bent crystal (qmBC) slab of a thickness $L$ and its orientation with respect to an incident beam of a transverse size $\sigma$ and angular divergence $\phi$. The point O denotes the center of the anticlastic curvature of radius $R_\mathrm{a}$. The Si(111) planes (thick dashed curves) are bent due to the QM effect with the radius of curvature $R_\mathrm{qm}$; $\theta_\mathrm{qm} = L/R_\mathrm{qm}$ represents the QM bending angle. $\theta_\mathrm{ex}$ is the deflection angle of a positron with respect to the beam direction. $\theta^\mathrm{ch.}_\mathrm{d}$ is the deflection angle of a channel}
\label{Figure1} 
	
\end{figure}

The silicon crystal under consideration, $Z_{Si} = 14$, has a diamond crystal structure and exhibits a `quasi-mosaic' (QM) bending of the (111) planes, see Figure \ref{Figure1}.

When two mechanical moments are applied to a crystal slab to obtain a primary curvature, a second curvature, known as anticlastic curvature, is also produced. Both curvatures act on the same planes, but in perpendicular directions, resulting in a saddle shape  \cite{Sushko_2022,Paulo_2024}.

Because of the periodic structure, crystals have an\-iso\-trop\-ic physical properties that give rise to a secondary curvature present in planes orthogonal to the primary curvature, known as QM bending.

The simulation parameters were determined on the basis of the experimental data  \cite{Mazzolari_2024}, where the specific value of the anticlastic radius ($R_\mathrm{a}$) was not provided. The simulations were performed using MBN Explorer  \cite{Solovyov_2012,Solovyov_2017,MBN_Explorer,Sushko_2019} for the `quasi-mosaic' Bent Crystal (qmBC) with parameters corresponding to the experimental ones  \cite{Mazzolari_2024}. The geometry of the crystal and its orientation with respect to the positron beam is schematically represented in Figure \ref{Figure1}. It is important to emphasize that this illustration is purely schematic and not drawn to scale. The radius of the anticlastic curvature, is on the order of meters, whereas the QM radius, denoted by  $R_\mathrm{qm}$, is on the order of centimeters. In contrast, the thickness of the crystal, denoted by $L$, is in the order of micrometers. Therefore, the actual curvatures are not as pronounced as shown in the figure. It is important to note that the deflection angle of a channel is equal to $\theta^\mathrm{ch.}_\mathrm{d}=\theta_\mathrm{qm}+\theta_\mathrm{e}$, where $\theta_\mathrm{qm}$ is the QM bending angle and $\theta_\mathrm{e}$ is the beam incidence  angle. The positron trajectory angle with respect to the channel centreline at the exit of the channel is the deflection of a positron with respect to the beam direction minus the channel deflection angle ($\theta_\mathrm{ex} - \theta^\mathrm{ch.}_\mathrm{d}$).

A range of $R_\mathrm{a}$  values from 10 to 100 m was explored in increments of 10 m. These precise choices were made carefully on the basis of the previous knowledge   \cite{Paulo_2024}. This analysis was extended to the larger anticlastic radii, and the limiting case $R_\mathrm{a} \to \infty$ m was investigated. This limiting case corresponds to a bent crystal (BC) with a single radius of curvature $R_\mathrm{qm}$.

The thickness of the crystal in the beam direction  was set to $L=29.9 \ \mu$m, with a QM bending radius of $R_\mathrm{qm} = 30.8\ $mm.
The simulations were performed with a Gaussian beam of positrons with an energy of $530$ MeV, a width  ($\sigma$) of $1.5\ $mm, and divergence ($\phi$) of $60\ \mu$rad, directed towards the crystal. The Lindhard angle \cite{Lindhard_1965}, denoted  as $\theta_L$, is approximately $300 \ \mu$rad. For comparison, simulations were also carried out with the beam divergences $\phi = 0, 20, 40$ and $60\ \mu$rad.

The beam was positioned at $y=h_0$, where $h_0$ is equal to the beam incidence angle $\theta_\mathrm{e} = 0 \ \mu$rad, resulting in parallel incidence to the bent planes of Si(111)  \cite{Sushko_2022,Paulo_2024}. In addition, simulations have been performed with $y=h$,  corresponding to the beam incidence angle $\theta_\mathrm{e}$ of $-75$, $-150$, $-300$, $-375$ and $-450\ \mu$rad, respectively. 

A total of $100,000$ trajectories were simulated for each set of parameters (anticlastic radius, divergence and beam entrance angle). The distribution of deflected positrons as a function of the deflection angle $\theta_\mathrm{ex}$ was plotted from $-1,000$ to $1,500\ \mu \mathrm{rad}$, i.e. in the same range as the experimental data. This range normally includes  $95\ \%$ of the simulated  trajectories.

The next sections are devoted to the presentation of the simulation results and the detailed analysis of the steering of  the $530$ MeV positron beam by the silicon qmBC described above. The role of the CH, DCH, VR, and VC processes in the steering effect at different orientations of the target crystals is elucidated by the trajectory analysis. This work continues the efforts started in  \cite{Mazzolari_2014,Wienands_2015,Sushko_2022,Paulo_2024,Mazzolari_2024} towards a detailed understanding of the effects caused by the interaction of ultra-relativistic electrons, positrons and other charged particles with qmBCs.

\section{Results and discussions}\label{Results}
\subsection{Angular distribution of deflected positrons. Comparison with experimental results}

Careful selection of the size, thickness, and curvature of the crystal is essential for reducing undesirable anticlastic deformation  \cite{Guidi_2009,Germogli_2015}. This deformation limits the effectiveness of the crystal in channeling ultra-relativistic charged particle beams. Research has shown that optimizing the  dimensions of the crystal can effectively suppress this deformation  \cite{Mazzolari_2014}.

Based on this, and considering that the anticlastic radii are often unknown, a series of simulations with different anticlastic radii  were performed for the qmBC, with the aim of obtaining a better agreement between the results of the simulations and the experiment.

\begin{figure*}
	\centering 
	\includegraphics[width = \textwidth]{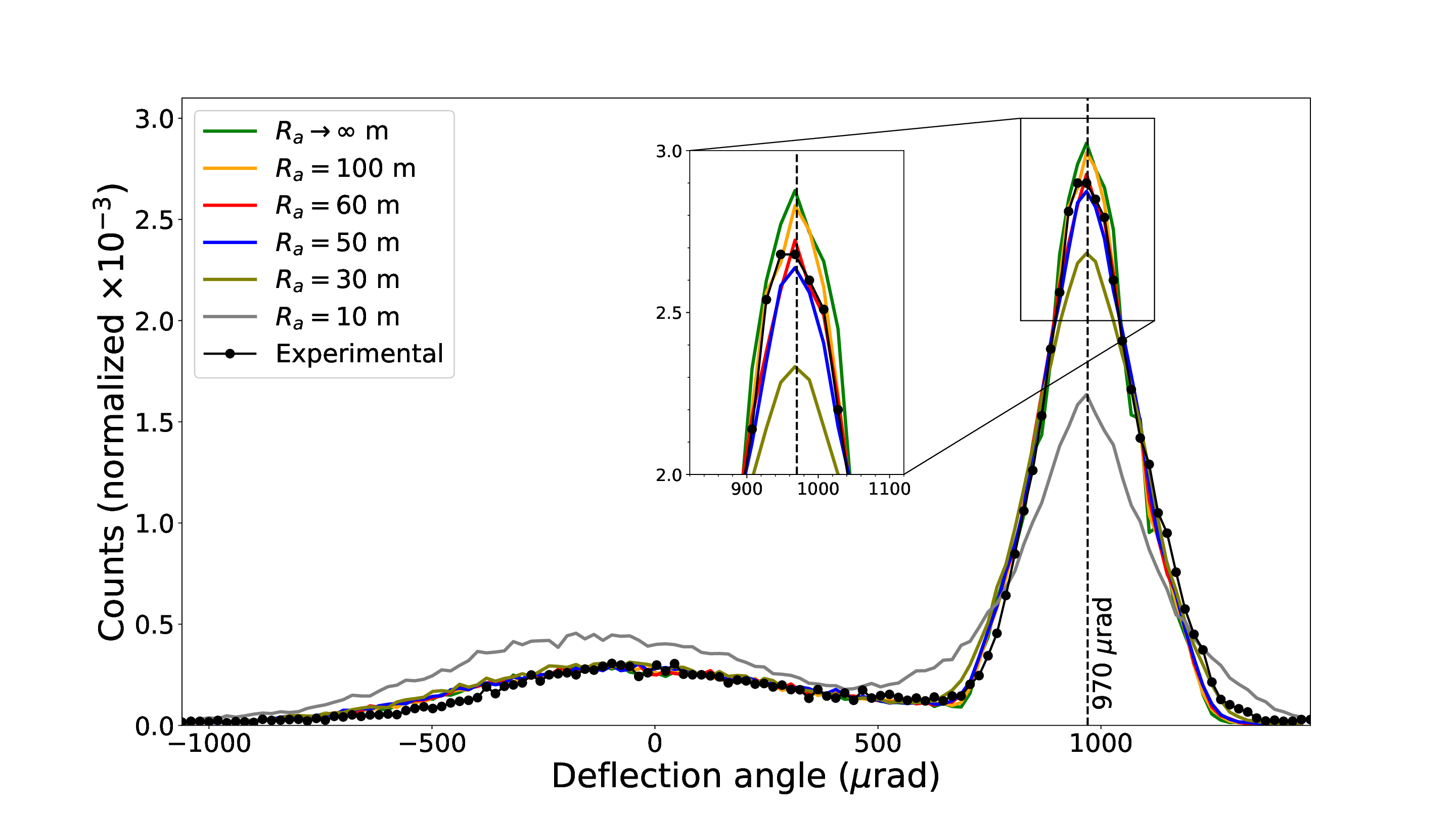}
	\caption{Angular distributions of positrons deflected by a Si(111) qmBC crystal at an angle of beam incidence $\theta_\mathrm{e}=0 \ \mu$rad. The continuous solid black circle line represents the experimental data \cite{Mazzolari_2024}, while the other lines show the results of the simulations for qmBC with  $R_\mathrm{a}$ = 10, 30, 50, 60 and 100 m, and the BC case} 
	\label{Figure2}
\end{figure*}

The distribution of the deflected positrons at the  Si(111) qmBC measured in the experiment for the angle of incidence of the beam  $\theta_\mathrm{e}=0 \ \mu$rad   is shown in Figure \ref{Figure2}. 
The angle $\theta_\mathrm{e}$ is defined as the angle between the direction of the incident beam center line at the crystal entrance and the bent QM plane. The solid black circles represent the experimental data  \cite{Mazzolari_2024}, while the other lines correspond to the results of simulations with different anticlastic radii ($R_\mathrm{a}=10, 30, 50, 60$ and $100\ $m). The green curve also represents the simulation result for the BC case. 

In the channeling regime, a significant part of the beam is deflected at the angle $970 \ \mu$rad, which is equal to the QM angle $\theta_\mathrm{qm}$. As reported in the experiments of Mazzolari et al.   \cite{Mazzolari_2024}, the left peak of the experimental curve is due to the deflection of particles that are not trapped in the channeling regime when they enter the crystal and thus move in the `over-barrier' regime. It is observed that this peak for $R_\mathrm{a}=10\ $m is shifted to the left compared to the other curves corresponding to the larger anticlastic radii studied. Specifically, the maximum is located at about $-103 \ \mu$rad. The other maxima arising for the larger anticlastic radii are located  between $-33$ and $-66 \ \mu$rad.

The closest agreement between the experimental results and the results of simulations is for $R_\mathrm{a}= 60\ $m. As the anticlastic radius decreases, the shape of the main peak in the distribution formed by the channeled posit\-rons changes. The height of the peak decreases as the width increases. This behavior is due to the  increase in the number of non-channeled  particles passing through the crystal as the anticlastic radius decreases. The increase in the height of the peak of the distribution of the non-channeled positrons and its shift to the left at $R_a=10\ $m is due to the enhancement of the volume reflection of positrons from the bent crystalline planes.

\begin{figure*}
	\centering 
	\includegraphics[width = \textwidth]{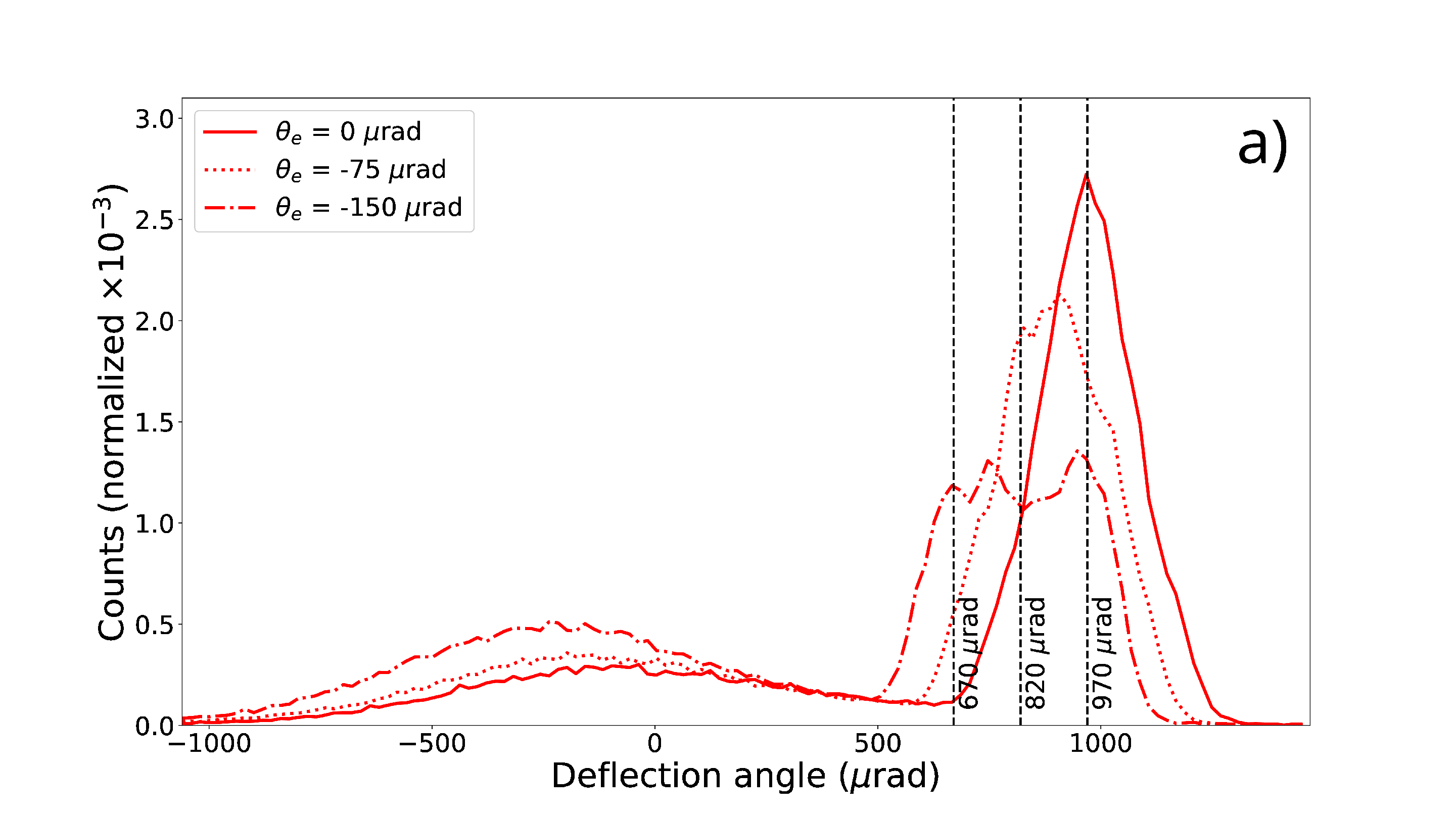}\\
	\includegraphics[width = \textwidth]{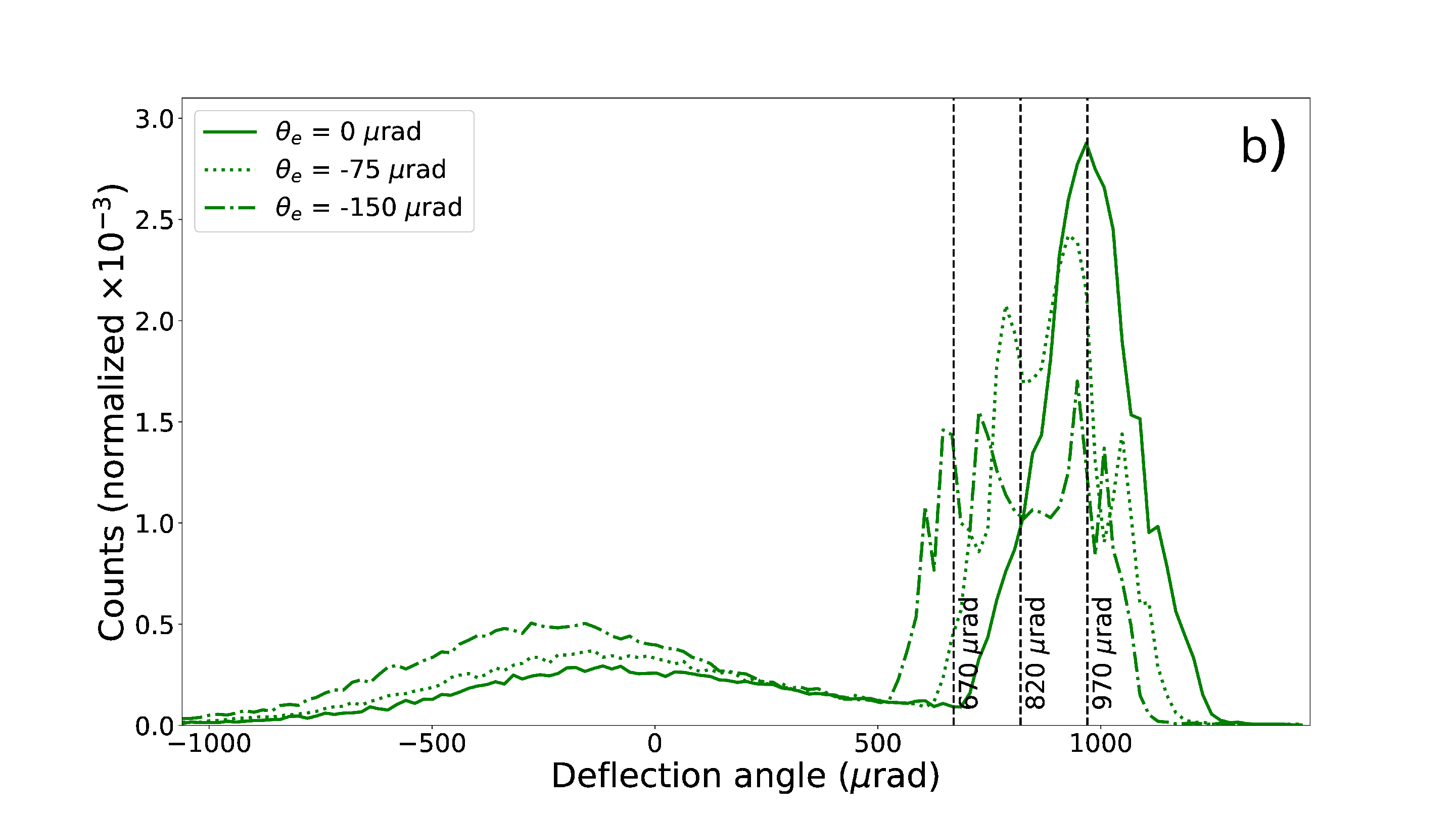}
	\caption{Variations in the angular distribution of deflected positrons at $\theta_\mathrm{e}=0 \ \mu$rad, $\theta_\mathrm{e}=-75 \ \mu$rad, and $\theta_\mathrm{e}=-150 \ \mu$rad for (a) qmBC with $R_\mathrm{a}=60\ $m and (b) BC case} 
	\label{Figure3}
\end{figure*}

To study the dependence of the angular distribution of the deflected positrons on the angle at which the beam hits the qmBC, simulations were carried out at different angles of incidence $\theta_\mathrm{e}$. The results of the simulations of the angular distributions of the deflected positrons at $\theta_\mathrm{e}=0 \ \mu$rad, $\theta_\mathrm{e}=-75 \ \mu$rad and $\theta_\mathrm{e}=-150 \ \mu$rad are shown in Figures \ref{Figure3}a (for $R_\mathrm{a}=60\ $m) and \ref{Figure3}b (for $R_\mathrm{a} \to \infty\ $m), where $\theta_\mathrm{e}$ is defined as in Figure \ref{Figure1}.  It can be seen that as the absolute value of $\theta_\mathrm{e}$ increases, the prominent  band of the distribution around $1,000 \ \mu$rad and below, formed by the channeled positrons, is shifted  towards the smaller deflection angles. This is easy to understand from the geometry of the system shown in Figure \ref{Figure1}. When the crystal is tilted positrons enter the channeling regime inside the crystal at some distance from its surface. This reduces the length that the positrons channel inside the crystal, and correspondingly reduces their angle of  deflection as they exit the crystal.  Figures \ref{Figure3}a and \ref{Figure3}b also show that the channeling peak in the distribution observed for $\theta_\mathrm{e}=0 \ \mu$rad decreases at  $\theta_\mathrm{e}=-75 \ \mu$rad and $\theta_\mathrm{e}=-150 \ \mu$rad and additional maxima appear. 

The  second prominent peak in the distribution is formed at negative deflection angles of positrons  which are not captured in the channeling regime and experience the volume reflection. This peak becomes higher and shifts towards \linebreak{} larger negative values of the deflection angles as the crystal is tilted towards the negative values $\theta_\mathrm{e}$ and the absolute values of $\theta_\mathrm{e}$ increase. Thus,  at  $R_\mathrm{a}=60 \ $m, these picks in the distribution are located at $-104 \ \mu$rad (at $\theta_\mathrm{e}=0 \ \mu$rad), $-165 \ \mu$rad (at $\theta_\mathrm{e}=-75 \ \mu$rad) and $-238 \ \mu$rad (at $\theta_\mathrm{e}=-150 \ \mu$rad). At  $R_\mathrm{a} \to \infty\ $m (BC case), these picks in the \linebreak{} distribution are located at $-111 \ \mu$rad (at $\theta_\mathrm{e}=0 \ \mu$rad), \linebreak{}  $-127 \ \mu$rad (at $\theta_\mathrm{e}=-75 \ \mu$rad ) and $-233 \ \mu$rad (at $\theta_\mathrm{e} = -150 \ \mu$rad).

Figures \ref{Figure3}a and \ref{Figure3}b show that there is a clear distinction between the positron distributions at $R_\mathrm{a}=60 \ $m and $R_\mathrm{a} \to \infty \ $m (BC case). In the case of BC, all the peaks in the distribution are more pronounced, while in the case of $R_\mathrm{a}=60$ m, some of them are smeared out.

The difference between the two cases in Figures \ref{Figure3}a and \ref{Figure3}b is due to the fact that for the infinite $R_\mathrm{a}$ (the BC case), the crystal surface is perpendicular to the beam direction. This allows the channels to accept more positrons under almost ideal conditions. For $R_\mathrm{a} \to 60\ $m the crystal surface is curved and the acceptance conditions are less ideal.

Based on the work of Sushko et al. \cite{Sushko_2022}, the maximum value of the range $\Delta h_\mathrm{max}$ with respect to $h_0$, within which the positrons can be accepted by the channels, 
is equal to $\Delta h_\mathrm{max}= \theta_L R_\mathrm{a}$. 
This means that
particles with a horizontal displacement $h$ falling within the interval $h_0 \pm \Delta h_\mathrm{max}$ will satisfy  the channeling condition. Thus, the total interval of the
horizontal distances within which positrons are effectively accepted in the channeling
regime is equal to $2 \theta_L R_\mathrm{a}$ and is centered around $h_0$. The probability of
accepting pos\-i\-trons into channels is highest at the point $h=h_0$, while it gradually decreases as the value of $\Delta h$ increases.

As $R_\mathrm{a} \to \infty\ $m, positrons enter the crystal over the entire width of the beam under the optimal conditions for acceptance by the channels. Therefore, the fine structure of the distribution of positrons deflected by their channeling in the bent crystal is more pronounced in this case.

\begin{figure*}
	\centering 
	\includegraphics[width = \textwidth]{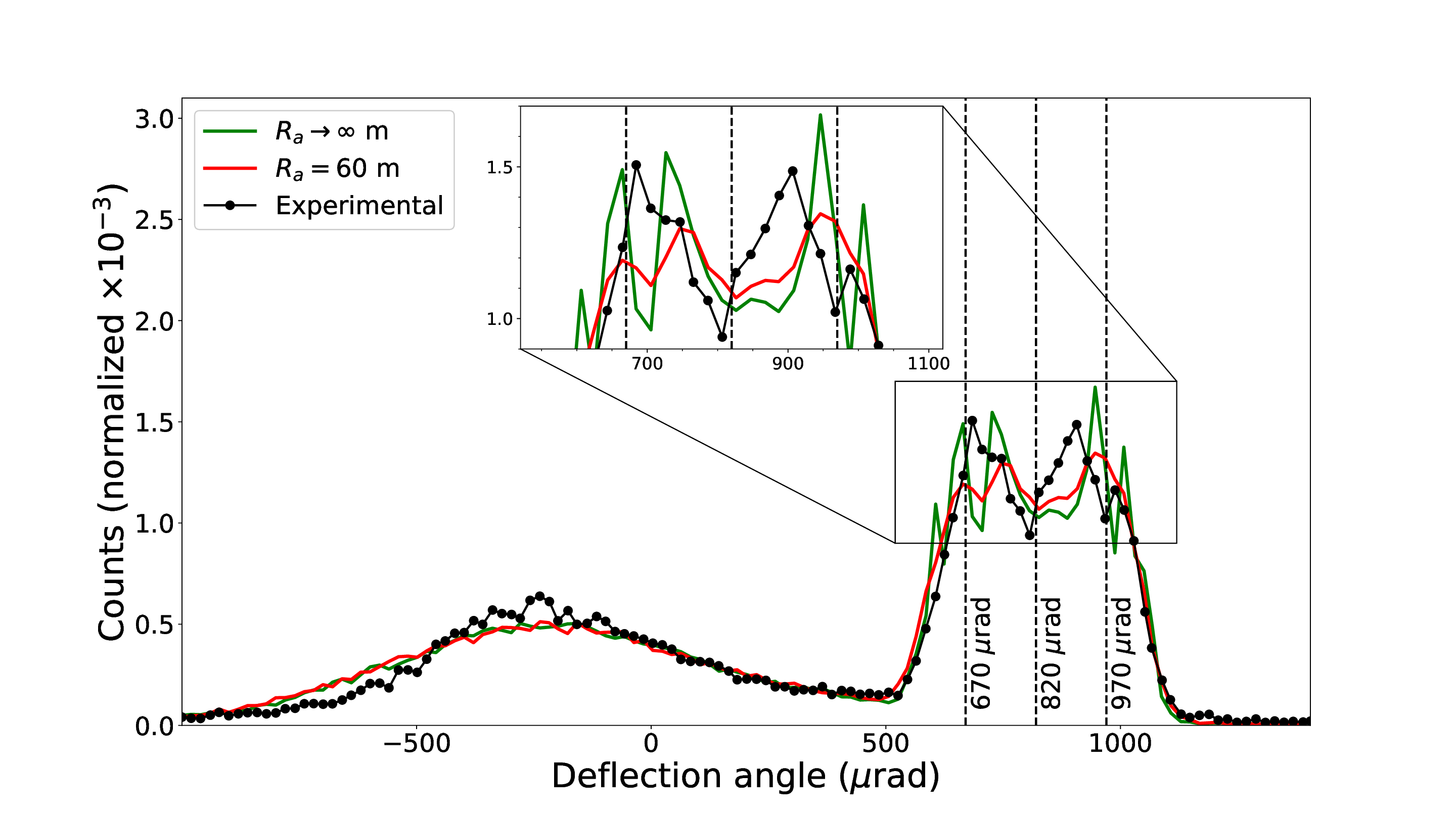}
	\caption{Experimentally measured angular distribution of positrons deflected by a Si(111) qmBC with the beam incidence angle of $-150 \ \mu$rad (black line). Simulated angular distributions of deflected positrons for  silicon (111) qmBC with $R_\mathrm{a}=60\ $m (red line) and $R_\mathrm{a} \to \infty\ $m(green line)} 
	\label{Figure4}
\end{figure*}

Experimental results  \cite{Mazzolari_2024} have shown the emergence of two distinct peaks within the channeling regime when the beam incidence  angle is set at $-150 \ \mu$rad. A comparative analysis of these experimental results alongside our simulations for $R_\mathrm{a}=60 \ $m and $R_\mathrm{a} \to \infty\ $m is shown in Figure \ref{Figure4}. The specific radius value of $R_\mathrm{a}=60\ $m was chosen because at this radius the best agreement of the simulation results with the experimental data was obtained for the beam incidence angle $\theta_\mathrm{e}=0 \ \mu$rad, as shown in Figure \ref{Figure2}. 

\begin{table}
	\begin{center}
	\caption{Normalized root mean square deviation (NRMSD in \%)  of the simulated angular distributions of the deflected positrons from the experimentally measured distribution}
	\begin{tabular}{|c|c|c|c|c|}\hline
		$\theta_\mathrm{e}$ & $R_\mathrm{a}$ & Non-channeling & Channeling & Total  \\ \hline
		$0 \ \mu$rad & $60\ $m & $3.32\ \%$ & $6.69 \ \%$ & $3.13 \ \%$  \\ \hline	
		$0 \ \mu$rad & $\infty\ $m & $3.13\ \% $ & $7.48\ \% $ & $3.24\ \% $  \\ \hline
		$-150 \ \mu$rad & $60\ $m & $4.09\ \% $ & $6.19 \ \%$ & $3.32\ \% $  \\ \hline
		$-150 \ \mu$rad & $\infty\ $m & $3.82\ \% $ & $6.36\ \% $ & $3.41\ \% $  \\ \hline
	\end{tabular}
	\label{Table1}
 	\end{center}	
\end{table}	

As can be seen in Figure \ref{Figure4}, the fine structure of the distributions for the channeled positrons derived in the simulations for both data sets is very close to that observed experimentally. However, a more detailed comparison shows that with $R_\mathrm{a}=60\ $m for both beam incidence angles ($\theta_\mathrm{e}=0 \ \mu$rad and $\theta_\mathrm{e}=-150\ \mu$rad) a better agreement with the experimental results is obtained. Table \ref{Table1} shows the normalised root mean square (RMS) deviations of the simulated angular distributions of the deflected positron beam from those measured in the experiment.

In contrast, for the non-channeled positrons, most of \linebreak{}which experience the volume reflection, the best agreement is found for $R_\mathrm{a} \to  \infty\ $m. However, when evaluating the whole distribution, the closest  match is also for $R_\mathrm{a}= 60 \ $m, as shown in Table \ref{Table1}.

\begin{figure*}
	\centering 
	\includegraphics[width = \textwidth]{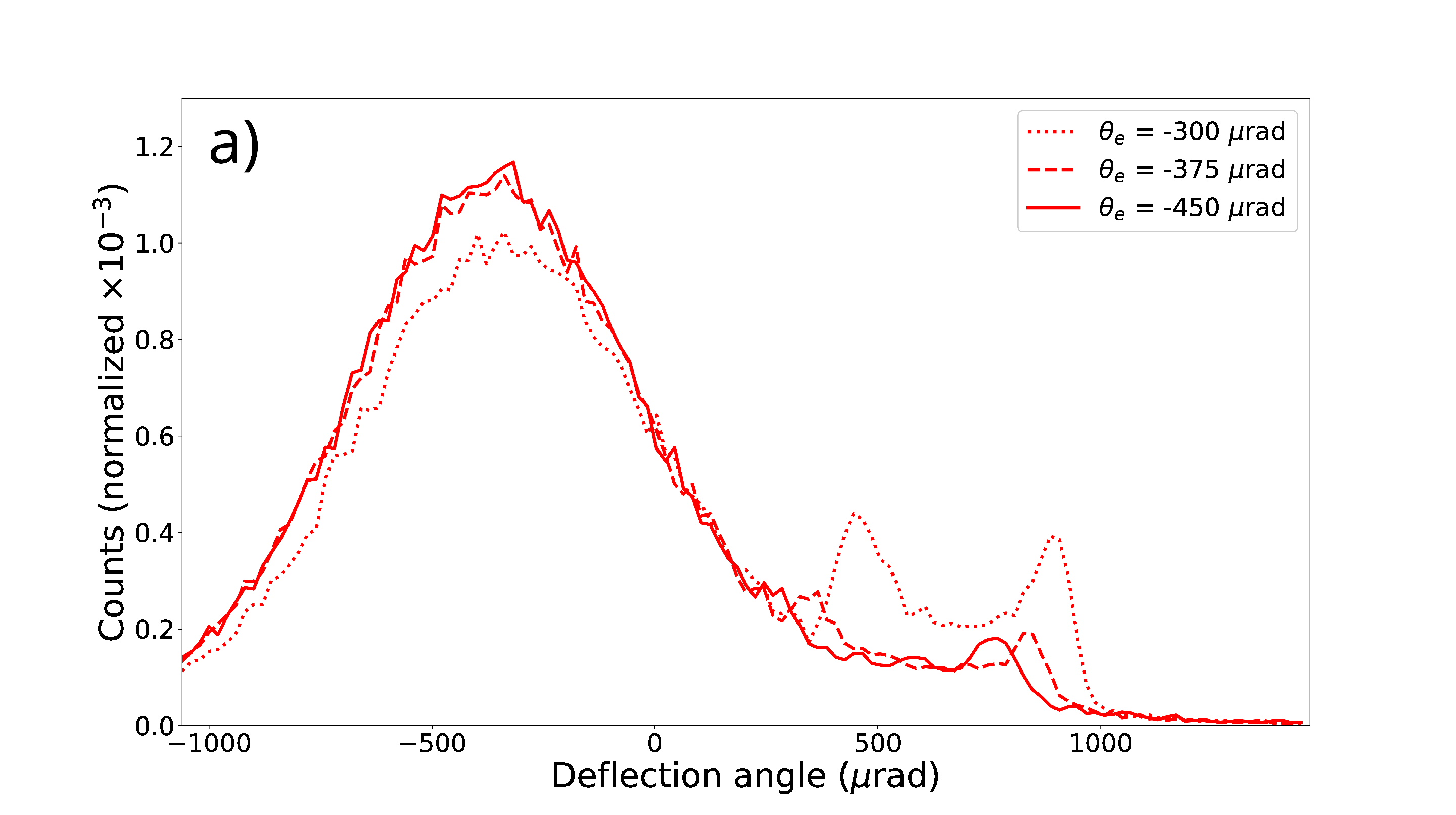}\\
	\includegraphics[width = \textwidth]{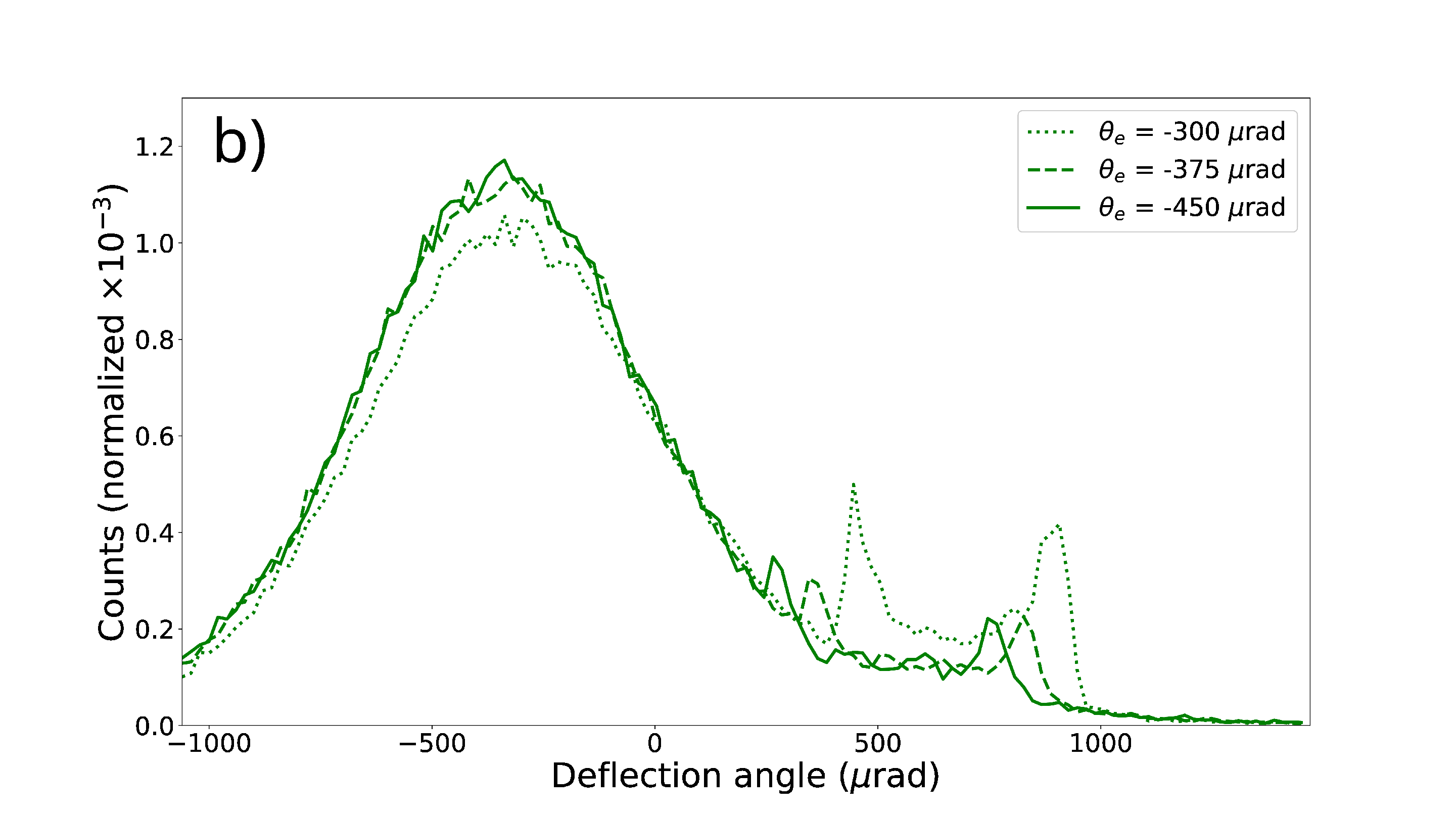}
	\caption{ Angular distribution of deflected positrons for the beam incidence  angles of $\theta_\mathrm{e}=-300 \ \mu$rad, $\theta_\mathrm{e}=-375 \ \mu$rad, and $\theta_\mathrm{e}=-450 \ \mu$rad for (a) qmBC with  $R_\mathrm{a}= 60\ $m and (b) BC} 
	\label{Figure5}
\end{figure*}

The angular distributions of the deflected positrons were meticulously studied by simulation to further investigate the phenomena of volume reflection (VR) and volume capture (VC) measured by experimentalists. This complex behavior is shown in Figure \ref{Figure5}.

The angular distribution of the deflected positrons was simulated for the beam incidence angles of $\theta_\mathrm{e}=-300 \ \mu$rad, $-375 \ \mu$rad, and $-450 \ \mu$rad for Si(111) qmBC with an anticlastic radius of $R_\mathrm{a}=60\ $m, as illustrated in Figure \ref{Figure5}a. For comparison, this simulation was repeated for $R_\mathrm{a} \to \infty\ $m. The result is shown in Figure \ref{Figure5}b. It is noteworthy that the results obtained for both anticlastic radii are similar. The small differences between the two cases can be explained by the same reasons as given above for the interpretation of the results shown in Figures \ref{Figure3}a and \ref{Figure3}b.

The result obtained for $\theta_\mathrm{e}=-300 \ \mu$rad suggests that for the beam incidence angle close to the Lindhard angle, a certain number of positrons are either directly accepted by the channels or undergo VC inside the crystal. This phenomenon is clearly demonstrated in our simulations. For Ra = $60\ $m, the number of channeled positrons is 10.61\%, while for $R_\mathrm{a} \to \infty \ \mathrm{m}$ it is $8.03\ \%$. 
In the simulation itself, a VC of $2.72\ \%$ was achieved for $R_\mathrm{a} = 60 \ $m and $2.75\ \%$ for the BC case.
	
For the beam incidence angles greater than the Lindhard angle (i.e. $\theta_\mathrm{e}=-375 \ \mu$rad, and $-450 \ \mu$rad), less than 2$\ \%$ of the positrons propagate through the crystal in the channeling regime. For such crystal orientations, a significant number of positrons (about 50$\ \%$) undergo VR.  See supplementary material for further information.

Figure \ref{Figure6}a shows the angular distributions of deflected positrons for different anticlastic radii at the beam incidence angle of $\theta_\mathrm{e} = -150\ \mu$rad. The left peak in the distribution corresponds to VR. The shape of the distributions shows a dependence on the anticlastic radius. For example, the VR peak is  at $-264 \ \mu$rad for $R_\mathrm{a}=10\ $m, while  for the larger anticlastic radii, the maximum values of the VR peak are in the range of values between $-230$ and $-238\ \mu$rad. Note also that the shape of the VR peak in the distribution does not change much for large values of the anticlastic radius $R_\mathrm{a} >30\ $m.

\begin{figure*}
	\centering 
	\includegraphics[width = \textwidth]{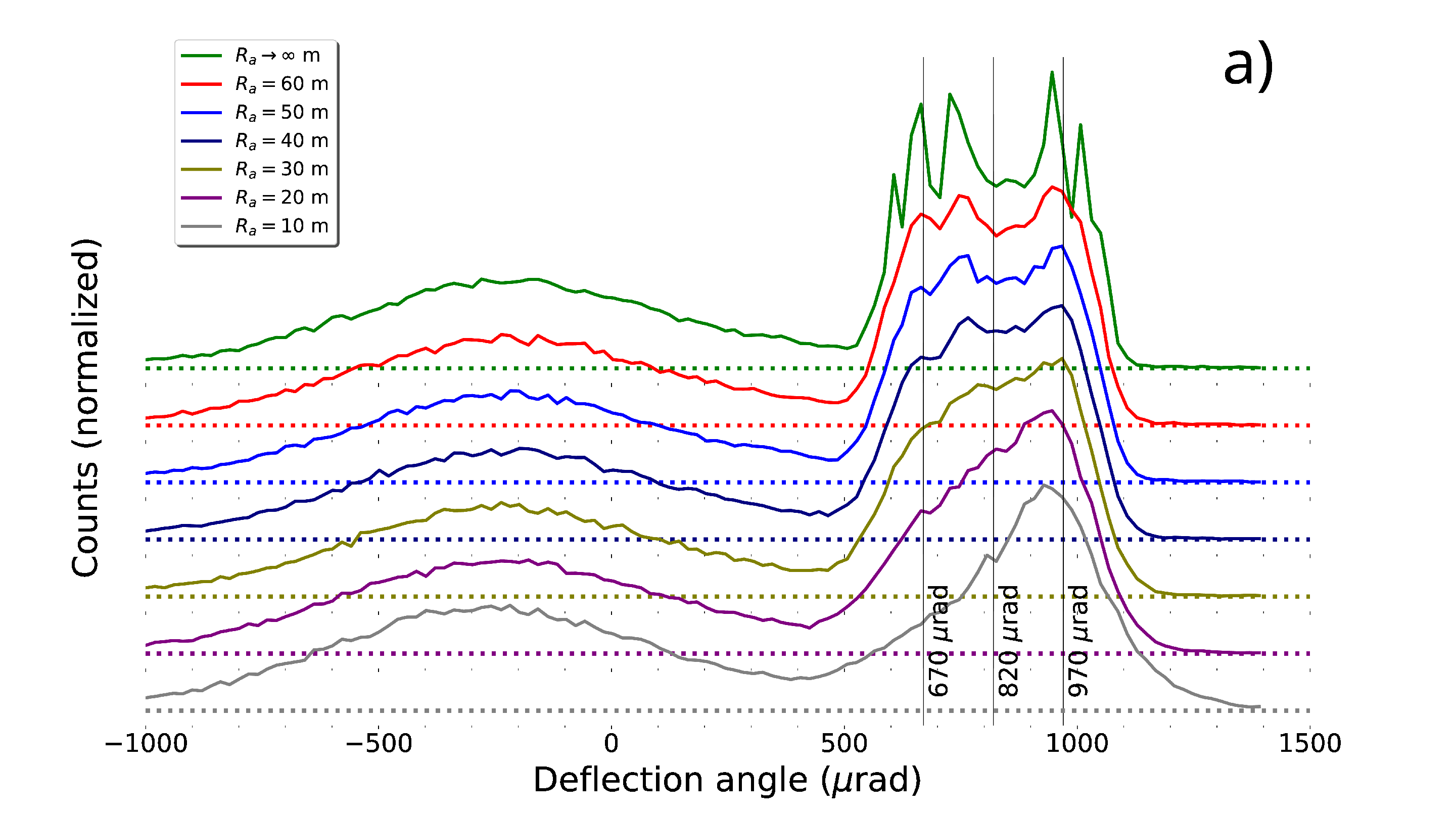}\\
	\includegraphics[width = \textwidth]{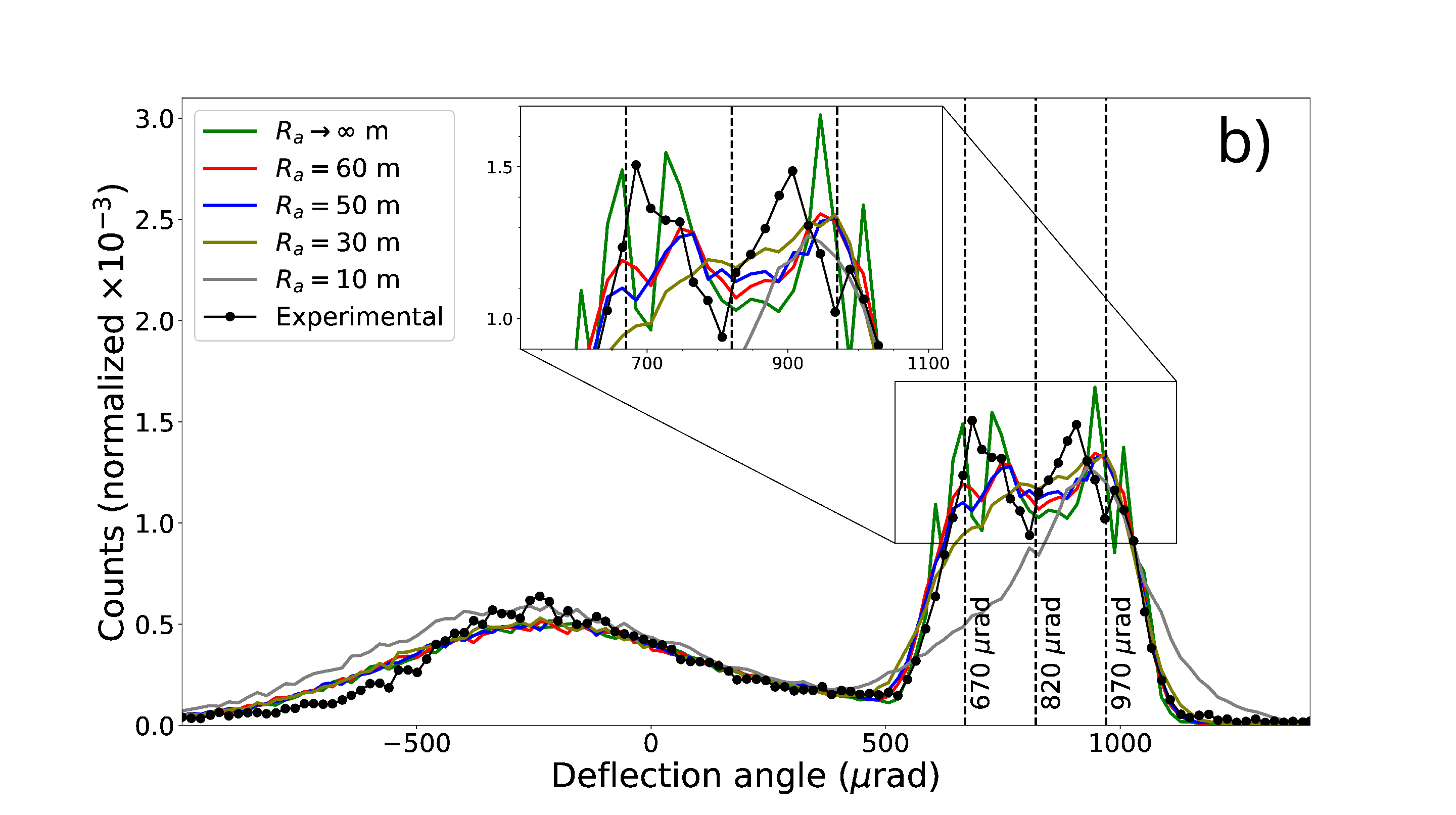}
	\caption{Angular distributions of deflected positrons for different anticlastic radii in qmBC with incidence angles of $\theta_\mathrm{e} = -150 \ \mu$rad. (a) $R_\mathrm{a}=10\ $m (gray line), $R_\mathrm{a}=20\ $m (violet line), $R_\mathrm{a}=30\ $m (olive line), $R_\mathrm{a}=40\ $m (dark blue line), $R_\mathrm{a}=50\ $m (blue line), $R_\mathrm{a}=60\ $m (red line), $R_\mathrm{a} \to \infty\ $m (green line). (b) The results shown in Figure 6a are grouped together and compared with the experimental result \cite{Mazzolari_2024} (solid circle black line)} 
	\label{Figure6}
\end{figure*}

It can be observed that for small anticlastic radii ($R_\mathrm{a}=10, 20,$ and $30\ \mathrm{m}$), the channeling band in the distribution has only one distinct peak. The shape of the channeling band changes significantly with increasing  anticlastic radius. For large anticlastic radii $R_\mathrm{a}=40, 50, 60\ $m and $R_\mathrm{a} \to \infty\ $m, two distinct peaks appear in the channeling band. This result is consistent with the findings of Mazzolari et al. \cite{Mazzolari_2024}. 

Figure \ref{Figure6}b groups together the results shown in Figure \ref{Figure6}a, corresponding to $R_\mathrm{a}=10, 30, 50, 60 \ $m, and $R_\mathrm{a} \to \infty\ $m, and compares them with the experimental results \cite{Mazzolari_2024}.

For the channeling band in the distribution, simulations performed with anticlastic radii of $50, 60\ $m and $R_\mathrm{a} \to \infty\ $m show close agreement with the experimental results. In particular, the result with $R_\mathrm{a}=60\ $m shows the closest agreement with the experimental data. This suggests that the fine structure of the channeling band is influenced by the anticlastic radius, as discussed above (see Figure \ref{Figure4} and Table \ref{Table1}).

Based on this analysis, it can be stated that the anticlastic radius of the Si(111) qmBC used in the experiment is approximately equal to $R_\mathrm{a} = 60 \ $m.

\subsection{Angular distribution of deflected positrons as a function of the angular divergence $\phi$ of the positron beam.}

Let us now consider the dependence of the angular distribution of the deflected positrons on the angular divergence of the incident beam, denoted by $\phi$.
To this end, the two values of $R_\mathrm{a}=60\ $m and $R_\mathrm{a} \to \infty\ $m were analyzed in the simulations.

\begin{figure*}
	\centering 
	\includegraphics[width = \textwidth]{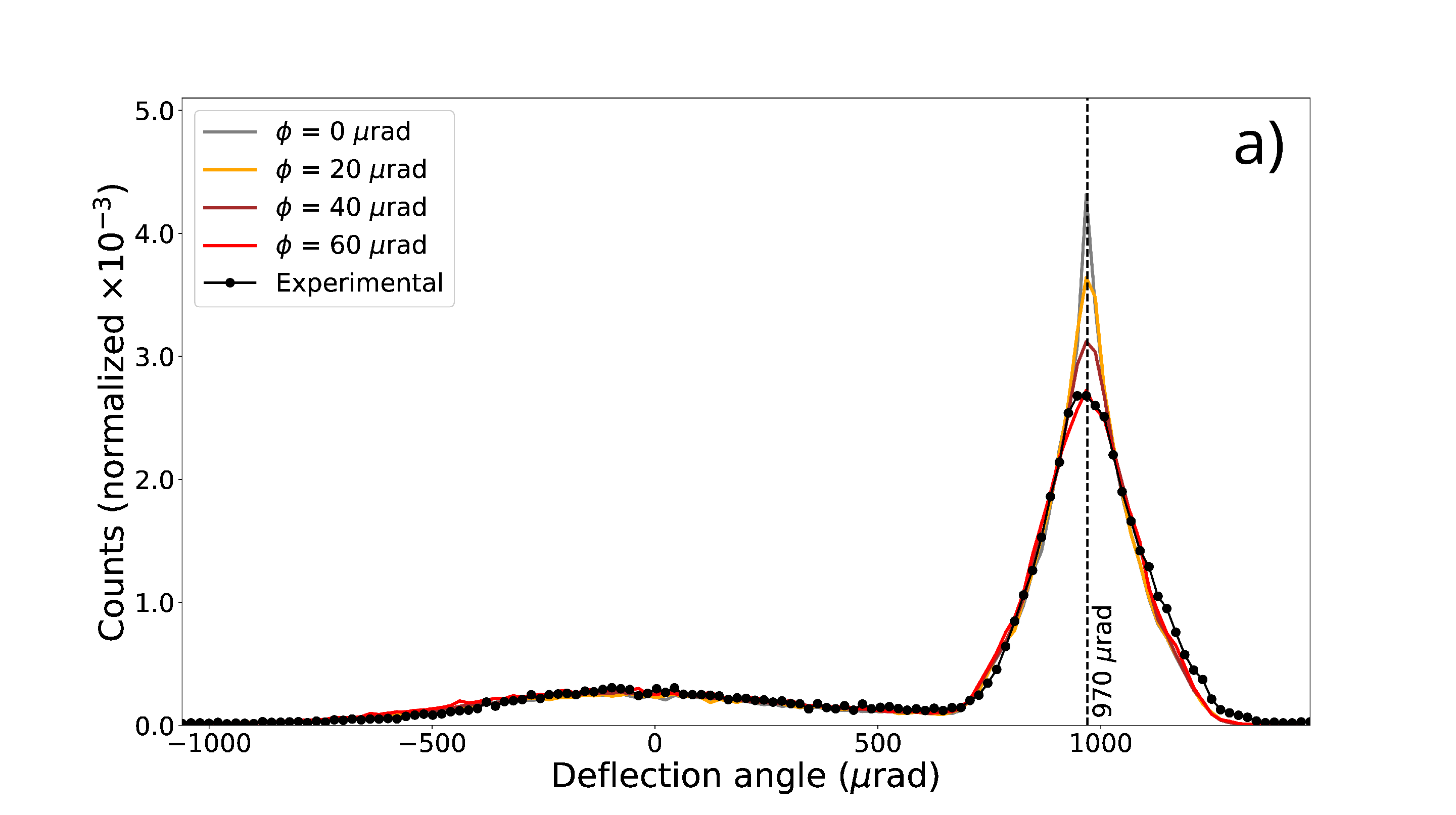}\\
	\includegraphics[width = \textwidth]{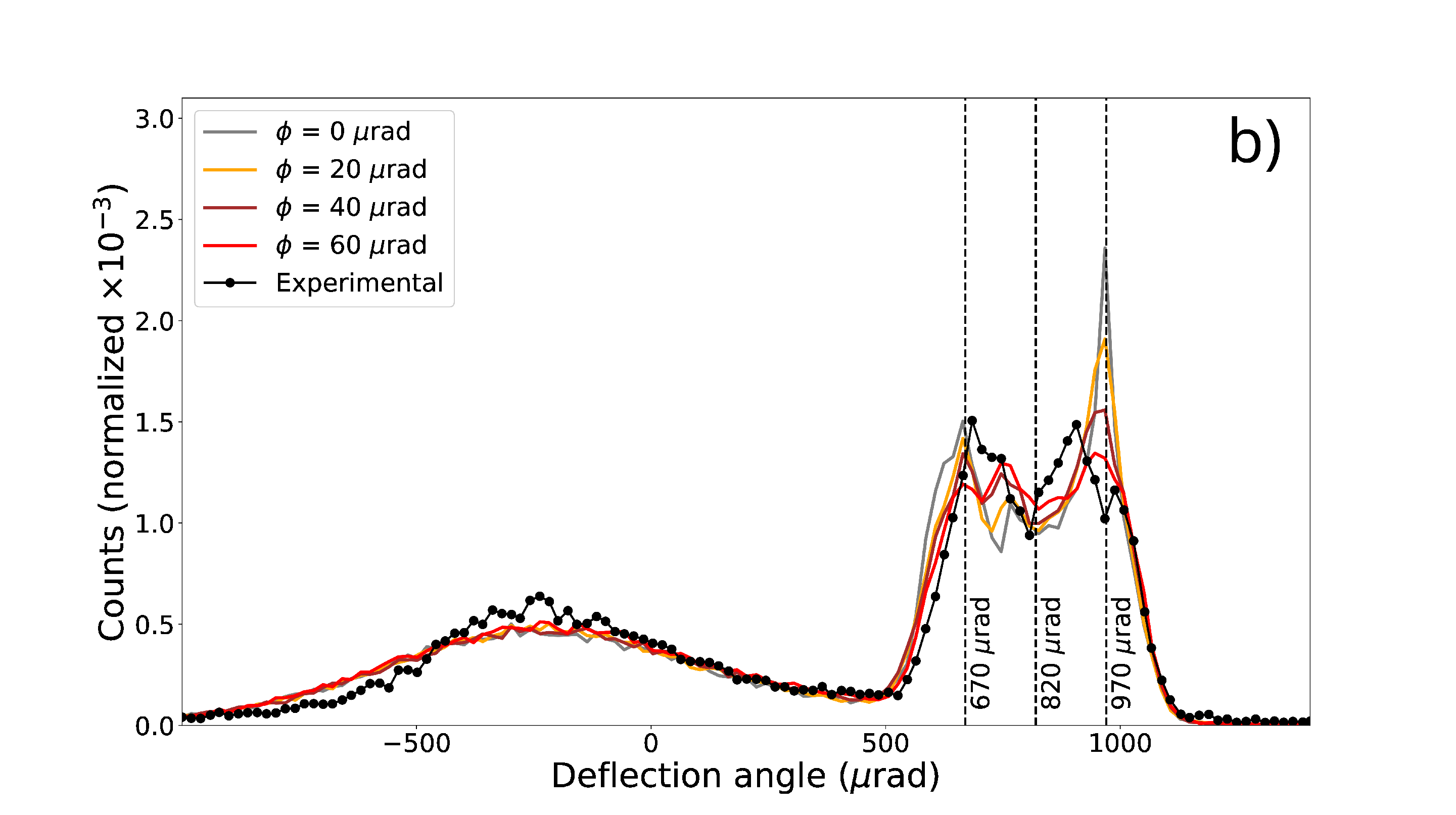}
	\caption{Angular distributions of deflected positrons as a function of  the divergence of the incident beam for an anticlastic radius of $R_\mathrm{a}=60\ $m. (a) Dependence for the beam incidence angle of $\theta_\mathrm{e}=0 \ \mu$rad. (b) Dependence for an beam incidence angle of $\theta_\mathrm{e}=-150 \ \mu$rad} 
	\label{Figure7}
\end{figure*}

Figure \ref{Figure7} shows the dependence of the angular distributions of the deflected positrons on the divergence of the incident beam for an anticlastic radius of $R_\mathrm{a}=60\ $m. Figure \ref{Figure7}a shows this dependence for the angle of incidence of the beam $\theta_\mathrm{e}=0 \ \mu$rad, while Figure \ref{Figure7}b shows the dependence for $\theta_\mathrm{e}=-150 \ \mu$rad. 

For both beam incidence angles, $\theta_\mathrm{e}=0 \ \mu$rad and \linebreak{}$\theta_\mathrm{e}=-150 \ \mu$rad, the best agreement with the experimental data was found for the beam divergence of $\phi=60 \ \mu$rad. This value is in agreement with the experimentally reported value.

\begin{figure*}
	\centering 
	\includegraphics[width =\textwidth]{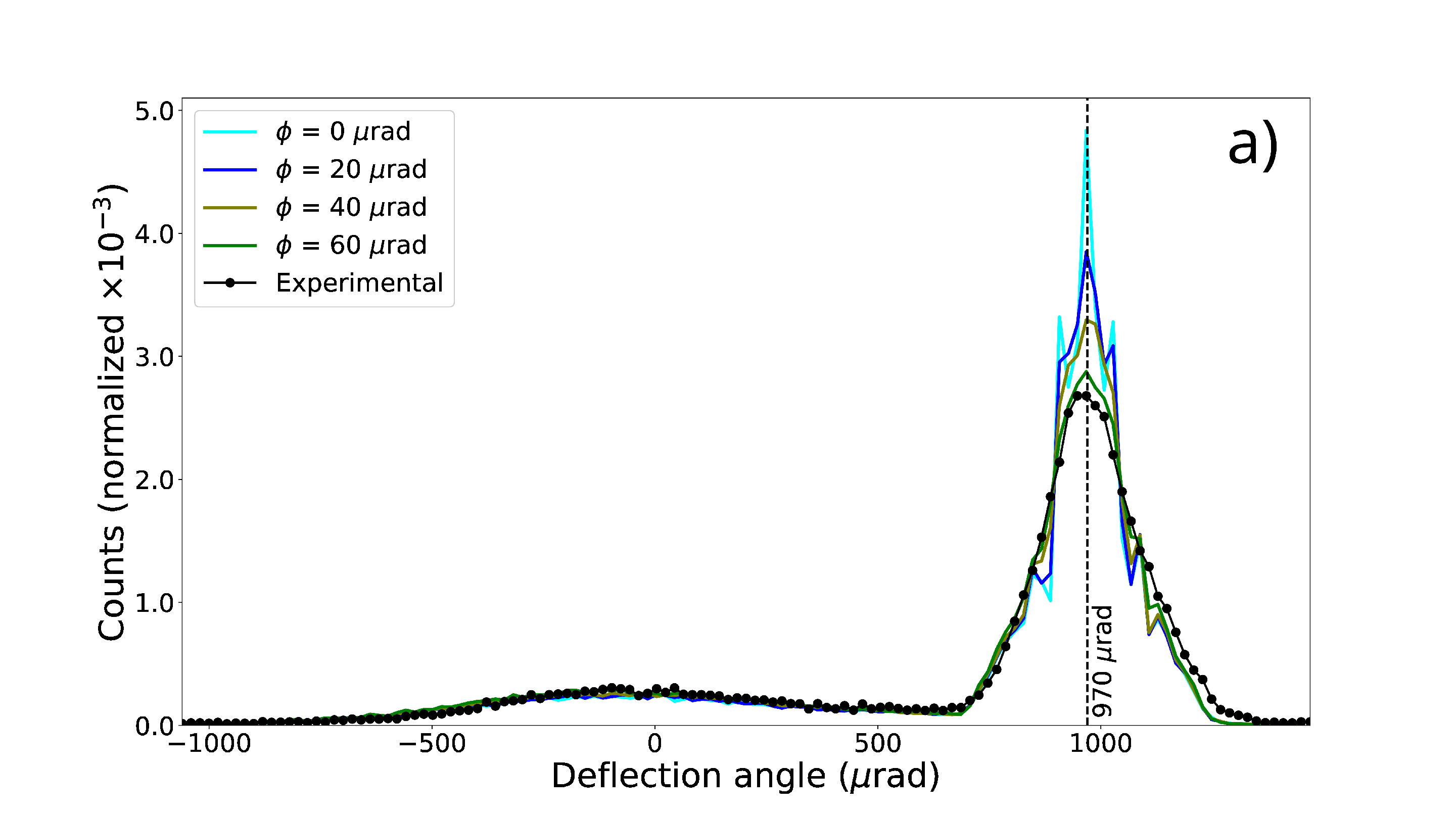}\\
	\includegraphics[width = \textwidth]{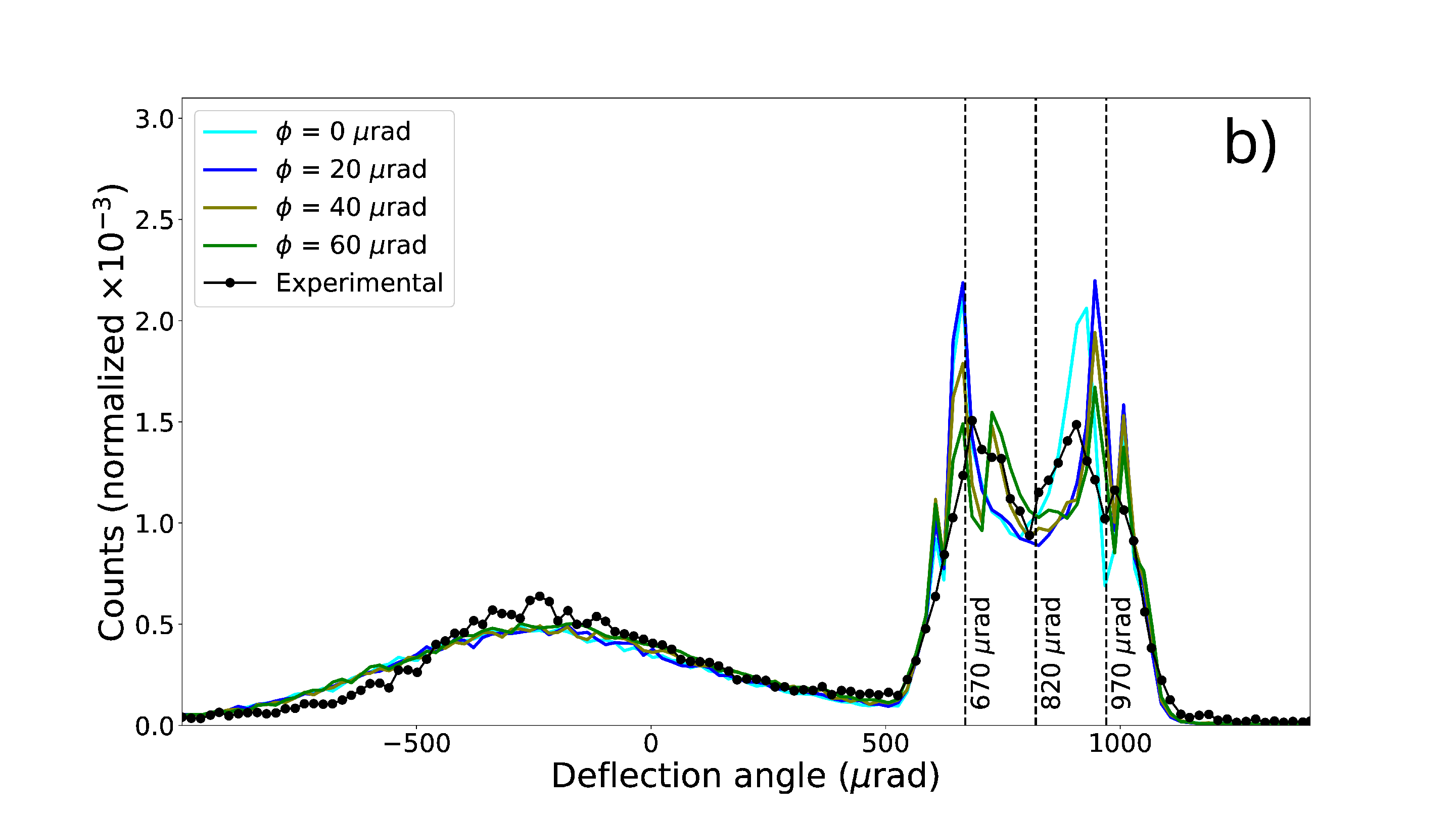}
	\caption{Angular distributions of deflected positrons  simulated for  different divergences of the incident beam with $R_\mathrm{a} \to \infty\ $m. 
(a) Dependence for an beam incidence angle of $\theta_\mathrm{e}=0 \ \mu$rad. (b) The same as in (a) for $\theta_\mathrm{e}=-150 \ \mu$rad} 
	\label{Figure8}
\end{figure*}

Figure \ref{Figure8} shows the dependence of the angular distributions of the deflected positrons on the divergence of the incident positron beam for $R_\mathrm{a} \to \infty\ $m. Figure \ref{Figure8}a shows the results for the angle $\theta_\mathrm{e}=0 \ \mu$rad, while Figure \ref{Figure8}b refers to $\theta_\mathrm{e}=-150 \ \mu$rad. The best correspondence between the experimental and simulated data is obtained with a beam divergence of $\phi=60 \ \mu$rad.

In both cases $R_\mathrm{a}= 60\ $m and $R_\mathrm{a} \to \infty\ $m, it can be seen  that the distributions of the deflected positrons become shar\-per with decreasing of the divergence of the incident beam. This result is in line with expectations, since a more narrowly focused beam should lead to a sharper angular distribution of the deflected positrons in the region of the channeling band, due to the increasing  number of positrons accepted by the crystal channels with decreasing  beam divergence.

As the divergence of the incident beam increases, so does the range of values of the angles of incidence of the positrons. As a result, the intensity of the channeling peak decreases. This can be observed at $R_\mathrm{a}=60\ \mathrm{m}$ and $R_\mathrm{a} \to \infty\ \mathrm{m}$, in Figures \ref{Figure7} and \ref{Figure8}. In Figure \ref{Figure8}, the fine structure in the channeling band of the distribution becomes sharper as in Figure \ref{Figure7}.

\subsection{Positron dynamics in the channeling regime}
Let us now analyze the origin of the fine structure of the channeling band in the distribution of the deflected pos\-i\-trons. As in the previous sections, this analysis is carried out for the  Si(111) qmBC. 

The transverse motion of a positron in the planar channeling regime is nearly sinusoidal. 
During the time interval $T$ it completes a full oscillation along the transverse direction of the bent channel. At the same time it moves forward  along the longitudinal direction by the distance $\lambda $. 
In the harmonic case, the period $T$ is the same for the transverse oscillations of particles with different amplitudes \cite{Korol_2014}. 
Since all the positrons move at nearly the same velocity in the longitudinal direction, the spatial period of the channeling oscillations $\lambda $ should also be nearly the same for all the trajectories in this case. 

The interaction of the positrons with the atomic planes in the Si(111) qmBC is similar, but not identical to harmonic.
Due to the presence of anharmonicity, the channeling trajectories can have different periodicities. This is in contrast to the purely harmonic interplanar potential, where they all have the same period. 
The deviation from the harmonicity of the positron motion results in different exit conditions of different groups of positrons from the crystal, leading to the variation of the channeling band of the angular distributions of the deflected positrons, as seen in  Figures \ref{Figure3}b, \ref{Figure4}, \ref{Figure6}, \ref{Figure7}b, and \ref{Figure8}b, and the formation of the specific peaks. Let us look at  this phenomenon in more detail.

The exit angle of the positron with respect to the channel deflection angle $\theta^\mathrm{ch.}_\mathrm{d}$ is determined by the phase $\varphi _{L}$ of the channeling oscillation at the exit of the channel and the amplitude $A$ of the channeling oscillations. For a sinusoidal trajectory, when the exit phase is $\varphi _{L} = 0$ or $\varphi _{L} = \pi$, the exit angle reaches an extremum.
The following analysis of the simulated positron trajectories shows that such points correspond to the positions of the peaks in the channeling band of the angular distribution of the deflected positrons.

The exit phase of a positron trajectory is equal to:
\begin{equation}
\varphi _{L} = \varphi _{0} +  2\pi \displaystyle \frac{L}{\lambda}, 
\label{eq3}
\end{equation}
where $\varphi _{0}$ is the initial phase at the channel entrance and the second term is the number of oscillations completed along the thickness $L$. The initial phase depends on the positron entrance point relative to the channel center line,  the positron angle of incidence $\theta_\mathrm{e}$ and the amplitude $A$ of the positron transverse oscillations. The presence of the different spatial periods in the positron motion discussed above allows the condition for an extremum to be satisfied  more than once at different $\lambda$, in contrast to the harmonic case, leading to the formation of several peaks in the channeling band of the angular distribution of the positrons.  

The positrons in the beam have different entrance conditions.
The entrance angle and the position of a positron at the channel entrance determine the parameters of the channeling oscillations (amplitude, spatial period and entrance phase). The oscillations along the thickness $L$ determine the characteristics of the positrons at the exit of the channel (exit phase and exit angle). 

For example, for a parallel incidence ($\theta_\mathrm{e} = 0\ \mu \mathrm{rad}$) the entrance phase of a sinusoidal trajectory is $\varphi _{0} = \pi /2$ or $\varphi _{0} = 3\pi /2$, depending on whether a positron enters a channel above or below its center.
In this case, the exit angles reach an extremum for the spatial periods satisfying the relation 
\begin{equation}
L/\lambda =1/4+k/2, 
\label{eq4}
\end{equation}
where $k$ is a positive integer number ($k=0, 1, 2, 3, ...$). These values of $\lambda$ determine the exit angles at which the peaks in the channeling band of the distribution of deflected positrons occur.

Let us divide all the trajectories of positrons channeled along the (111) planes in Si(111) qmBC into three groups according to the values of their deflection angles \( \theta_\mathrm{ex} \).
The first  group contains trajectories that are nearly aligned with the direction of the bent Si(111) channels at the exit point, i.e. \( \theta_\mathrm{ex} \approx \theta^\mathrm{ch.}_\mathrm{d} \). 
The second group includes trajectories with \( \theta_\mathrm{ex} < \theta^\mathrm{ch.}_\mathrm{d} \), while the third group includes those with \( \theta_\mathrm{ex} > \theta^\mathrm{ch.}_\mathrm{d} \). 

\begin{figure*}
	\centering 
	\includegraphics[width =\textwidth]{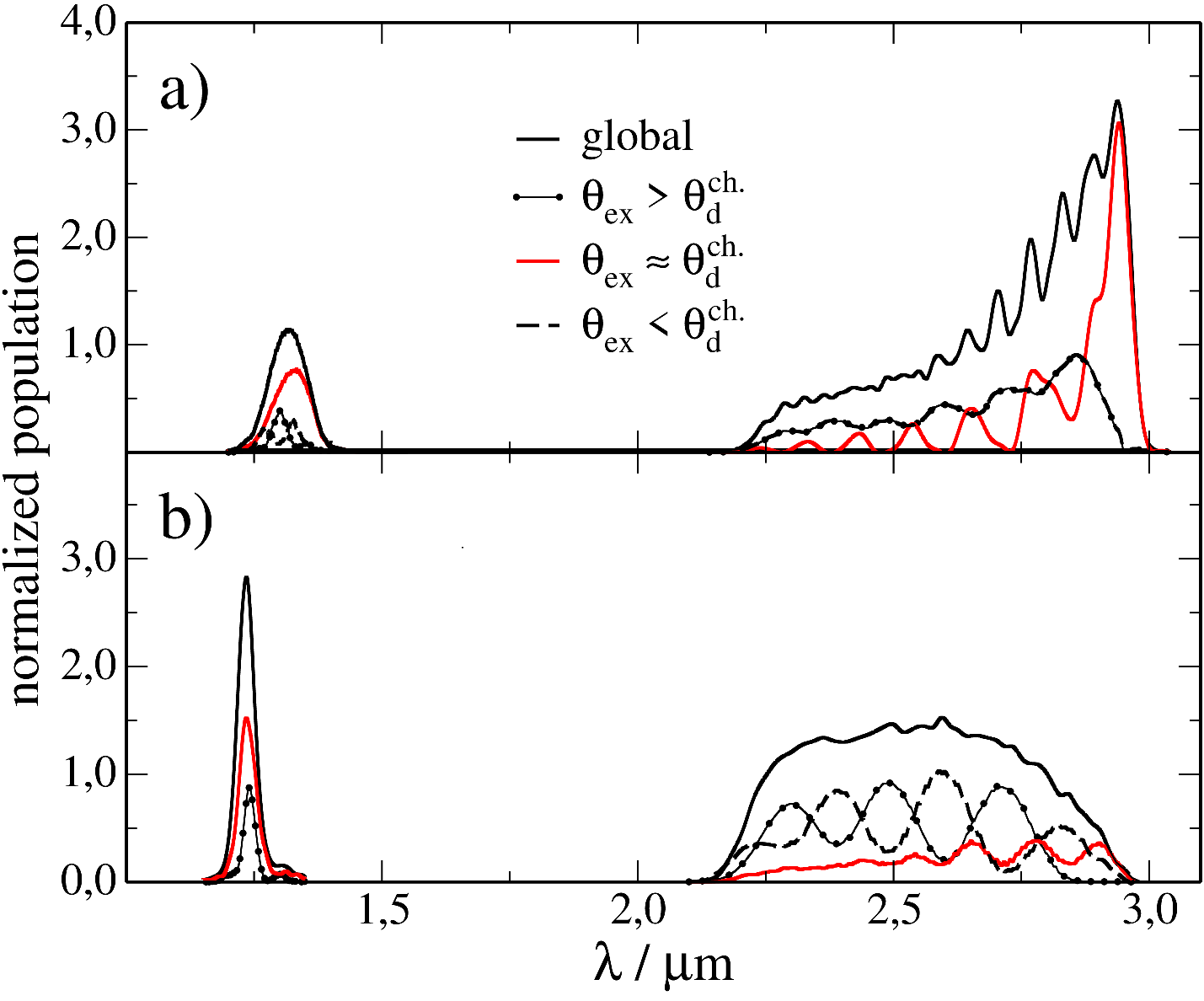}\\
	\caption{Population of the spatial periods of channeling oscillations for different groups of positron trajectories (see text for the definition) channeled through a $29.9\ $mm thick  Si(111) qmBC. For beam incidence angle of $0 \ \mu$rad (panel a) and $-150 \ \mu$rad (panel b). The values of $\theta^\mathrm{ch.}_\mathrm{d}$ are equal to $970 \ \mu$rad (panel a) and $820 \ \mu$rad (panel b).The beam divergence is equal to $60\ \mu$rad} 
	\label{Figure9}
\end{figure*}

The normalised histogram for the population of different groups of trajectories with respect to the spatial periods of the channeling oscillations is shown in Figure \ref{Figure9}.  The top panel of this figure corresponds to the beam incidence angles of $\theta_\mathrm{e} = 0\ \mu \mathrm{rad}$, while the bottom panel shows the result for $\theta_\mathrm{e} = -150\ \mu \mathrm{rad}$. On the left side of the figure (in both panels), the three maxima in the range of $\lambda$ from 1.1 $\mu$m to 1.4 $\mu$m correspond to the three groups of positron trajectories discussed above.   These maxima show the contributions of the groups of the positron trajectories passing the crystal through the so-called narrow channels in Si(111) crystals (see supplementary material for further information). The positions of the three maxima are very close to each other. In this case the population of the trajectory group with $\theta_\mathrm{ex} \approx \theta^\mathrm{ch.}_\mathrm{d}$ is the largest.

On the right side of Figure \ref{Figure9} (in both panels) the populations of the different trajectory groups for positrons passing the crystal in the channeling regime through the so-called wide Si(111) channels are shown as a function of length $\lambda$ introduced above. In this case the spatial period range of $\lambda$ is  from $2.0 \ \mu$m to $3.0 \ \mu$m. 
These curves show that for $\theta_\mathrm{e} = 0\ \mu \mathrm{rad}$ (panel a) the group of trajectories with $\theta_\mathrm{ex} \approx \theta^\mathrm{ch.}_\mathrm{d}$ oscillates and becomes dominant in the distribution of the trajectory populations in several regions of $\lambda$.  For  $\theta_\mathrm{e} = -150\ \mu \mathrm{rad}$ (panel b)  the situation is reversed, the groups of trajectories with $\theta_\mathrm{ex} < \theta^\mathrm{ch.}_\mathrm{d} $ and $\theta_\mathrm{ex} > \theta^\mathrm{ch.}_\mathrm{d} $ become dominant. Furthermore, the populations of these two groups oscillate as a function of $\lambda$, and each of them becomes dominant for certain values of $\lambda$.

\begin{figure*}
	\centering 
	\includegraphics[width =\textwidth]{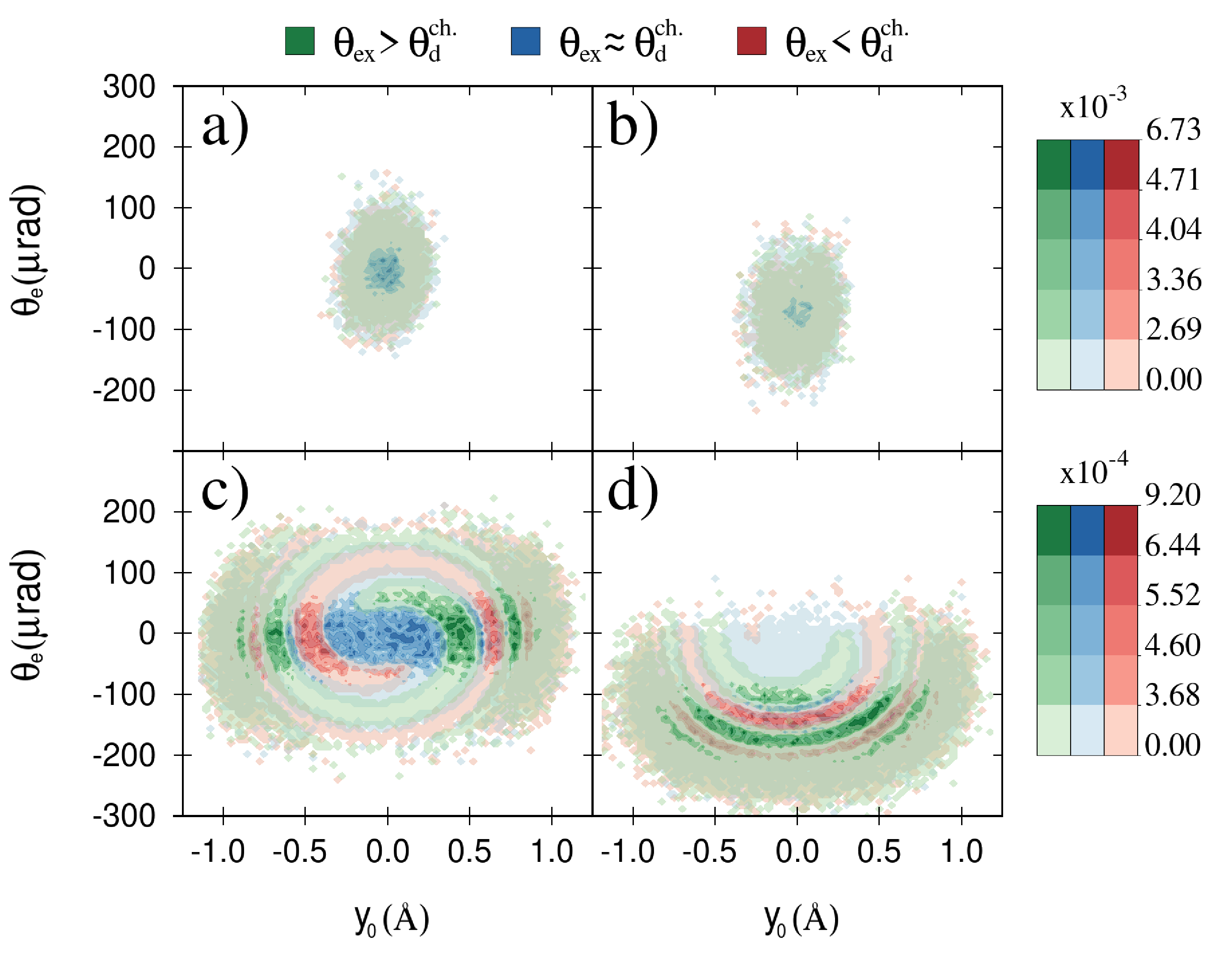}\\
	\caption{Correlation between the initial conditions ($y_0$, $\theta_\mathrm{e}$) and the $\theta_\mathrm{ex}$ for qmBC with an anticlastic radius $R_\mathrm{a}=60 \ $m. Panels a) and c) show the results for the narrow and the wide channels, respectively, at the beam incidence angle $\theta_\mathrm{e} = 0 \ \mu$rad. Panels b) and d) show the results for the narrow and wide channels, respectively, at the beam incidence angle $\theta_\mathrm{e} = -150 \ \mu$rad. In all panels, each point ($y_0$, $\theta_\mathrm{e}$) has been assigned a color indicating the positron trajectory group to which the point belongs. The color intensity indicates the population of the points. 
The beam divergence is equal to $60\ \mu$rad} 
	\label{Figure10}
\end{figure*}

To better understand such a behavior, let us now examine the correlation between the initial conditions for 530 MeV  positrons at the entrance of a bent Si(111) channel and their deflection angles $\theta_\mathrm{ex}$  after channeling through the crystal. Each point in the initial configuration space (defined by the variables $y$ and $\theta_\mathrm{e}$) has been assigned a color indicating the positron trajectory group to which the point belongs. This initial configuration space is shown in Figure \ref{Figure10} for a $29.9 \ \mu$m thick Si(111) qmBC with $R_\mathrm{a}= 60 \ $m and beam divergence of $60 \ \mu$rad for both channels and for two beam incidence angles ($\theta_\mathrm{e}=0 \ \mu$rad and $\theta_\mathrm{e}=-150 \ \mu$rad).

Panel a) of Figure \ref{Figure10} shows that the largest population of positrons steered through narrow channels arises at the direction of the channels ($\theta_\mathrm{ex} \approx \theta^\mathrm{ch.}_\mathrm{d}$ ). The initial configuration space of trajectories shows a symmetric distribution for all three groups of positrons with different $\theta_\mathrm{ex}$. The Lindhard critical angle for the Si(111) narrow channel is smaller than that for the wide Si(111) channel and is approximately equal to  $190 \ \mu$rad. This explains the absence of data points for positron incidence angles greater than $200 \ \mu$rad. Panel b) shows the same, but for the beam incidence angle $\theta_\mathrm{e}=-150 \ \mu$rad.

Interesting results are shown in panels c) and d) of the Figure \ref{Figure10} for wide Si(111) channels at beam incidence angles of $0 \ \mu$rad and $-150 \ \mu$rad, respectively. Panel c) shows a pattern resembling three spirals. Two of these spirals (red and green) extend from the central region shown in blue in panel c) towards the periphery, while the third spiral is located in the central region and leads to an extension of its borders. Each point on the green or red spiral represents initial conditions in the configuration space for the groups of positron trajectories with $\theta_\mathrm{ex}>\theta^\mathrm{ch.}_\mathrm{d}$ or $\theta_\mathrm{ex}< \theta^\mathrm{ch.}_\mathrm{d}$ respectively. The central region  of the initial configuration space (shown in blue) represents the group of positron trajectories with $\theta_\mathrm{ex} \approx \theta^\mathrm{ch.}_\mathrm{d}$.

Panel c) shows that the main contributions to the different groups of positron trajectories come from the range of  positron incident angles between $-50 \ \mu$rad and $50 \ \mu$rad, with the largest populations of trajectories having $\theta_\mathrm{ex} \approx \theta^\mathrm{ch.}_\mathrm{d}$. Table \ref{Table2} shows the statistical contributions of each trajectory group for both wide and narrow channels at the two beam incidence angles, based on the analysis of $60,000$ numerically simulated channeled trajectories of $530$ MeV positrons.

Panel d) shows,  in contrast to c), that the central region of the initial configuration space is almost depopulated at the beam incidence angle $\theta_\mathrm{e}=-150 \ \mu$rad. The main contributions for each group of positron trajectories come from in the lower spiral arms and the populations of the positron trajectories with $\theta_\mathrm{ex}>\theta^\mathrm{ch.}_\mathrm{d}$ or $\theta_\mathrm{ex}< \theta^\mathrm{ch.}_\mathrm{d}$ dominate.

\begin{figure*}
	\centering 
	\includegraphics[width =\textwidth]{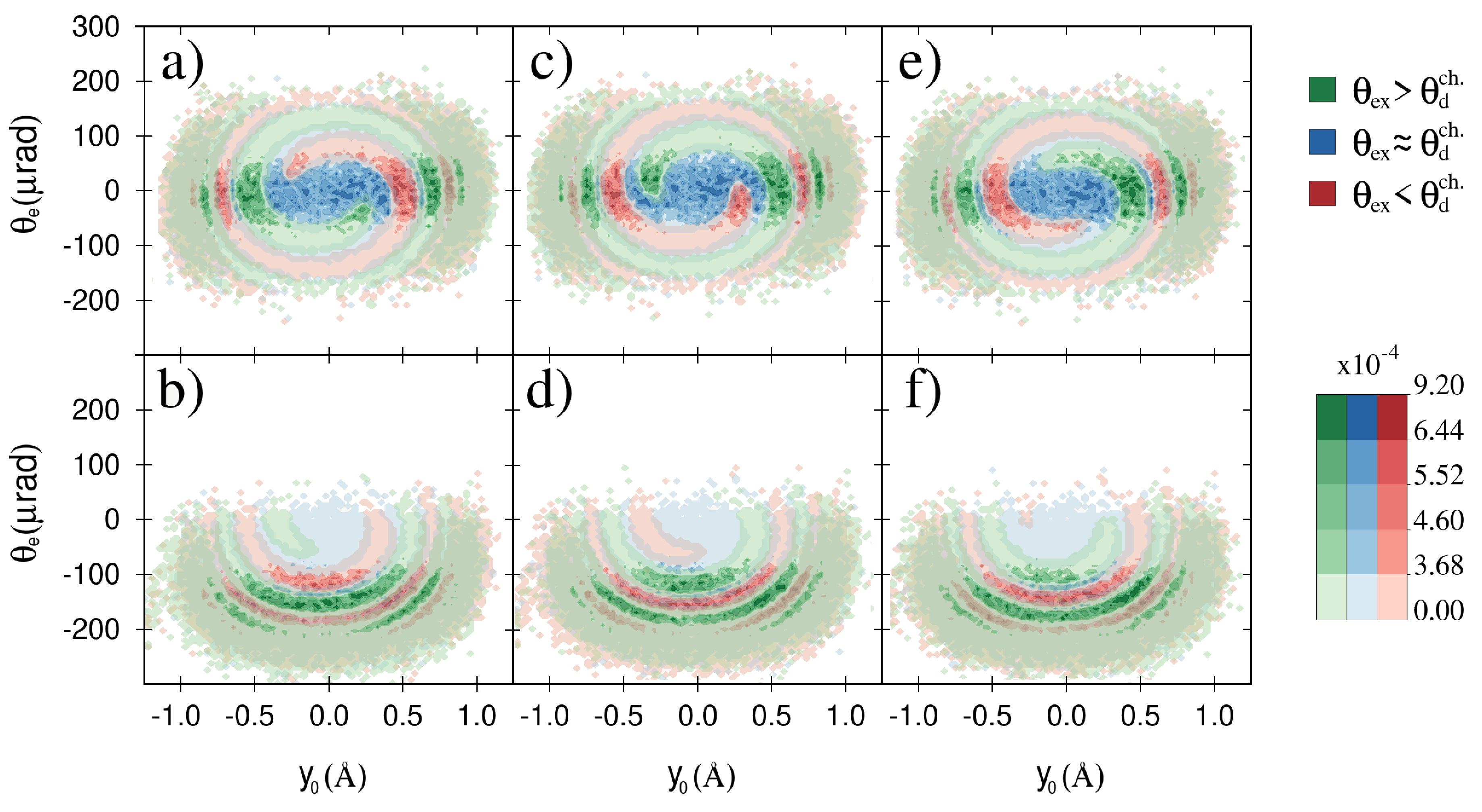}\\
	\caption{Correlation between the initial conditions $(y_0, \theta_\mathrm{e})$ and $\theta_\mathrm{ex}$. Panels a), c) and e) correspond to a beam incidence angle $\theta_\mathrm{e} = 0 \ \mu$rad and a crystal thickness of $29.9 \ \mu$m, $29.5 \ \mu$m and $29.1 \ \mu$m respectively. Panels b), d) and f) correspond to an beam incidence angle $\theta_\mathrm{e}=-150 \ \mu$rad and the same crystal thicknesses. In all panels, each point ($y_0$, $\theta_\mathrm{e}$) has been assigned a color indicating the positron trajectory group to which the point belongs. The color intensity indicates the population of the points. The beam divergence is equal to $60\ \mu$rad} 
	\label{Figure11}
\end{figure*}

The correlations shown in panels c) and d) of Figure \ref{Figure10} for a $29.9 \ \mu$m thick crystal are also present for other crystal thicknesses. At an angle of beam incidence of $0 \ \mu$rad, variations in $L$ cause rotations in the pattern observed in the initial configuration space. However, the population of positron trajectories with $\theta_\mathrm{ex} \approx \theta^\mathrm{ch.}_\mathrm{d}$  dominates in all cases shown in Figure \ref{Figure11} in panels a), c), and e).

The populations of the positron trajectory groups in the initial configuration space show a similar behavior for the angle of incidence of the beam $-150 \ \mu$rad. The population of the initial configuration space by different groups of positron trajectories changes with the variation of the crystal thickness; however, the populations of the groups with $\theta_\mathrm{ex}>\theta^\mathrm{ch.}_\mathrm{d}$ or $\theta_\mathrm{ex}< \theta^\mathrm{ch.}_\mathrm{d}$ remain dominant, see  panels b), d), and f) in Figure \ref{Figure11}.

\begin{table}
	\centering
	\caption{Population of different groups of positron trajectories (see definition in the text) passing through the so-called wide and narrow channels of a $29.9\ \mu$m thick  Si(111) qmBC. The beam divergence is equal to $\phi=60 \ \mu$rad. The beam incidence angles are $\theta_\mathrm{e}=0 \ \mu$rad and $\theta_\mathrm{e}=-150 \ \mu$rad}
	\begin{tabular}{|c|c|c|c|c|}\hline
	 & $\theta_\mathrm{e}$&$\theta_\mathrm{ex} \approx \theta^\mathrm{ch.}_\mathrm{d}$ & $\theta_\mathrm{ex} > \theta^\mathrm{ch.}_\mathrm{d}$  & $\theta_\mathrm{ex} < \theta^\mathrm{ch.}_\mathrm{d}$  \\ \hline
	 \multirow{2}{35pt}
	{\centering Narrow channel} & $0\ \mu \mathrm{rad}$    & $66\ \%$ & $17\ \%$ & $17\ \%$  \\ \cline{2-5}
	 & $-150\ \mu \mathrm{rad}$ & $55\ \% $ & $22\ \%$ & $23\ \%$  \\ \hline	
	 \multirow{2}{35pt}
	{\centering Wide channel} & $0\ \mu \mathrm{rad}$      & $36\ \% $ & $32\ \%$ & $32\ \%$  \\ \cline{2-5}
	 & $-150\ \mu \mathrm{rad}$ & $18\ \% $ & $40\ \%$ & $42\ \%$  \\ \hline
	\end{tabular}
	\label{Table2}
\end{table}

\begin{figure*}
	\centering 
	\includegraphics[width=\textwidth]{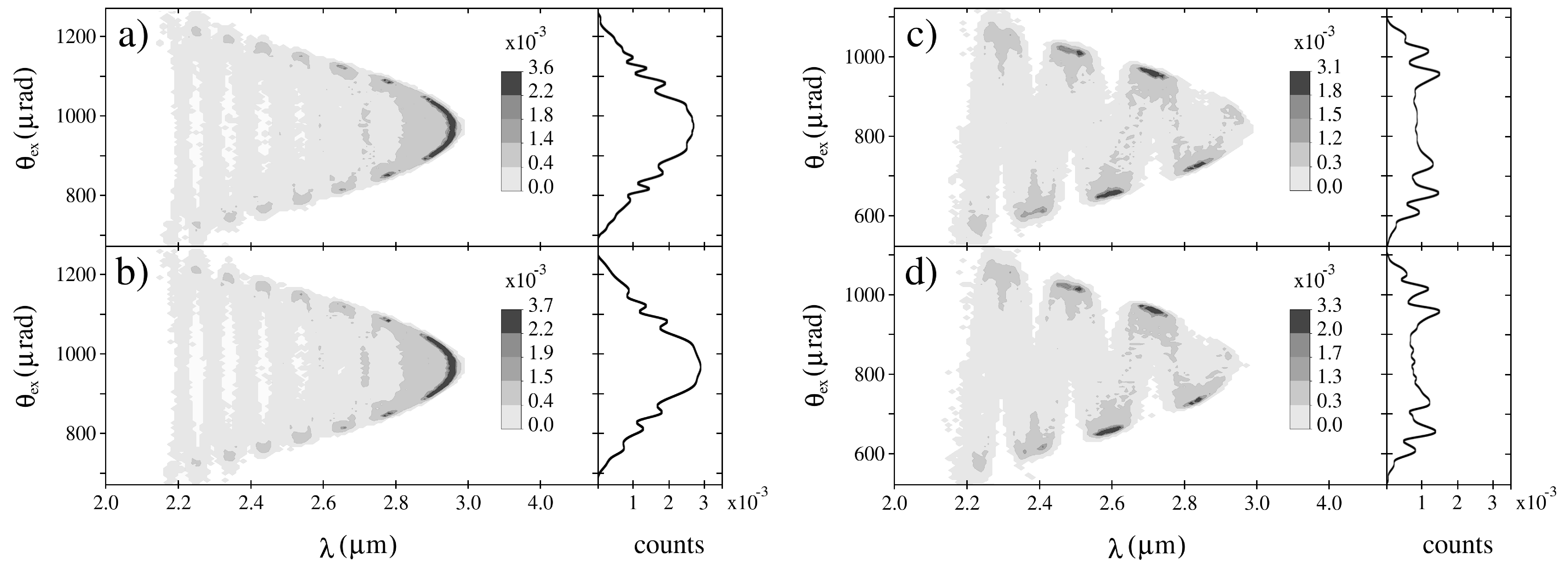}\\
	\caption{Correlations between spatial period of channeling oscillations and deflection angle. Panels a) and b): Beam incidence angle of $\theta_\mathrm{e}=0 \ \mu$rad. Panels c) and d): Beam incidence angle of $\theta_\mathrm{e}=-150 \ \mu$rad. Panels a) and c) correspond to qmBC with $R_\mathrm{a}=60\ $m, while panels b) and d) correspond to BC. In all cases, the beam divergence is $\varphi=60 \ \mu$rad} 
	\label{Figure12}
\end{figure*}

Figure \ref{Figure12} shows the distribution of positrons behind the crystal as a function of  the deflection angle $\theta_\mathrm{e}$ and as a function of the spatial period of the channeling oscillations $\lambda$. Panels a) and b) are for a positron beam centered at an  angle of incidence $\theta_\mathrm{e} = 0\ \mu \mathrm{rad}$ for the qmBC with the anticlastic radius of $R_\mathrm{a} = 60\ \mathrm{m}$ and $R_\mathrm{a} \rightarrow \infty\ \mathrm{m}$, respectively. Similarly, panels c) and d) are for an angle of incidence of $\theta_\mathrm{e} = -150\ \mu \mathrm{rad}$. The darker regions in the plot are  the regions of $\lambda$  where the phase of the positron trajectories at the exit of the crystal, see Eq. (\ref{eq3}), satisfies the extremum condition discussed above, see  Eq. (\ref{eq4}). Note that for $\theta_\mathrm{e} = 0\ \mu \mathrm{rad}$ there is symmetry in the distribution with respect to the deflection angle of a channel $\theta^\mathrm{ch.}_\mathrm{d}$, see Figure \ref{Figure1}. This occurs because the spatial periods $\lambda$ satisfying the extremum condition, see Eq. (\ref{eq4}), are the same for both groups of positron trajectories with $\theta_\mathrm{ex}>0$ and $\theta_\mathrm{ex}<0$. This is a result of the same entrance phases, see Eq.(\ref{eq3}), for these two groups of positron trajectories. The phase becomes a maximum for one group of trajectories at the values of $\lambda$ at which the phase is a minimum for the other group, and vice versa.  In the case of a non-zero angle of incidence  (e.g. $\theta_\mathrm{e} = -150\ \mu \mathrm{rad}$) the entrance phase for the two groups of trajectories takes different values and the symmetry in the dynamics of the two groups is lost.

\begin{figure}
	\centering 
	\includegraphics[width=\linewidth]{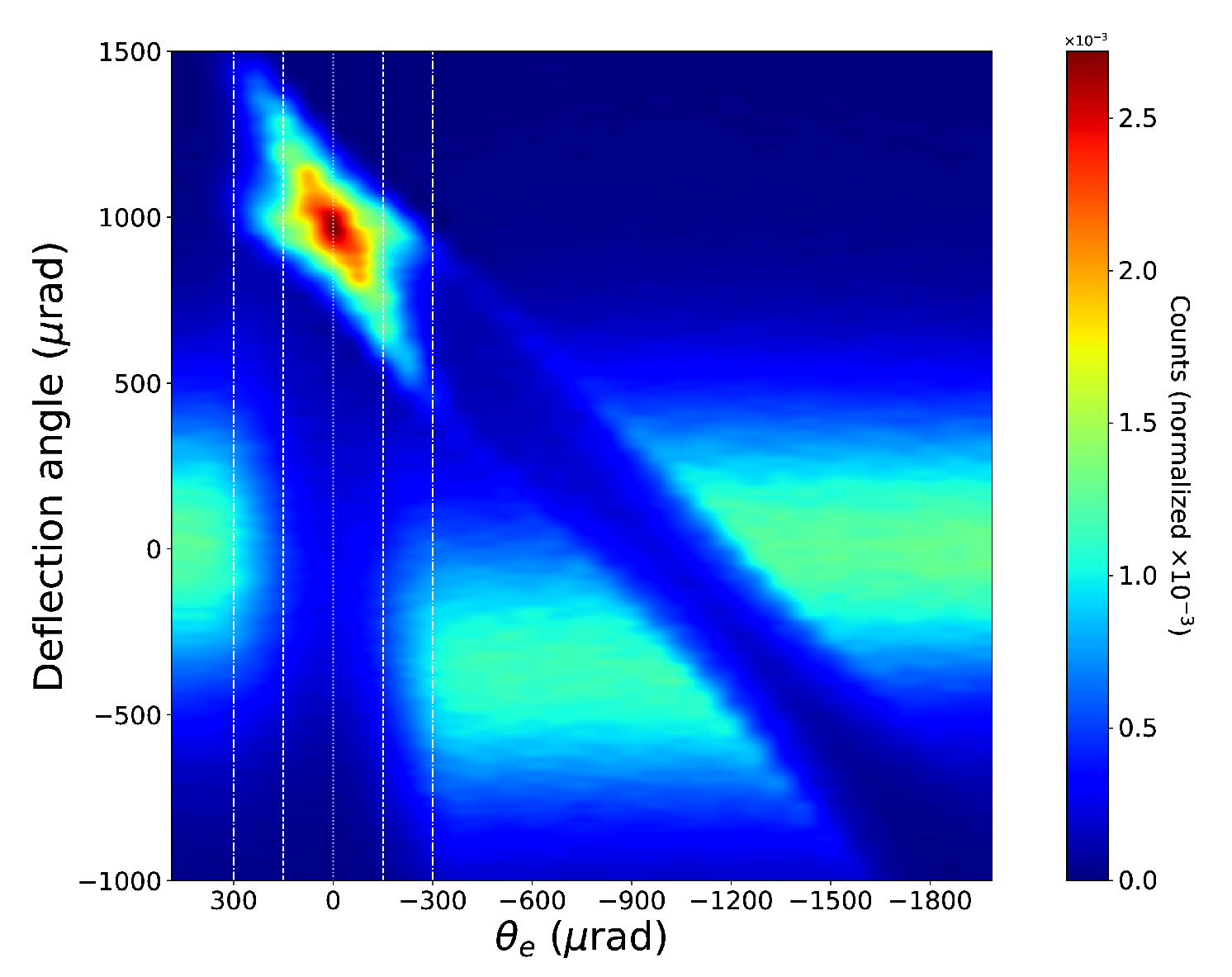}\\
	\caption{ Angular scan of the distribution of deflected positrons as a function of the beam incidence angle}
	\label{Figure13}
\end{figure}

Figure \ref{Figure13} shows an angular scan of the distribution of deflected positrons as a function of the beam incidence angle. The beam incidence angles range from $-1,950$ to 450 $\mu \mathrm{rad}$, with increments of 75 $\mu \mathrm{rad}$. The positron deflection angle ranges from $-1,000$ to $1,500$ $\mu \mathrm{rad}$.  All the configurations were simulated with an anticlastic radius of $R_\mathrm{a} = 60\ \mathrm{m}$ and a beam divergence of 60 $\mu \mathrm{rad}$. The data are plotted with a two-dimensional histogram of 127$\times$33 intervals, using two-dimensional cubic splines to obtain a smooth plot.

The regions associated with the different processes can be clearly identified:
\begin{itemize}
	\item{Over-barrier region: deflection angle around 0 $\mu \mathrm{rad}$, \linebreak{}beam incidence angle from 0 to 300 $\mu \mathrm{rad}$ and less than -1,000 $\mu \mathrm{rad}$;}
	\item{Volume reflection: deflection angle from -600 to -200 $\mu \mathrm{rad}$, beam incidence angle from -1,000 to -300 $\mu \mathrm{rad}$;}
	\item{Channeling: deflection angle from 500 to $1,500\ \mu \mathrm{rad}$, beam incidence angle from -300 to 300 $\mu \mathrm{rad}$.}
	
\end{itemize}

\section{Conclusions}\label{Conclusion}
This work provides the further evidence that  the relativistic molecular dynamics approach implemented in the advanced multi-purpose MBN Explorer allows an accurate computational modelling of the propagation of ultra-relativistic \linebreak{}projectiles in crystalline media. This atomistic approach allowed us to simulate the distribution of deflected positrons when a Si(111) qmBC is exposed to a 530 MeV positron beam. 

These simulations were carried out for Si(111) qmBCs with different anticlastic radii from 10 to 100 m and for a BC which is a limiting case of qmBC with an infinite anticlastic radius.  The angle of incidence of the beam on the crystals of -150 $\mu \mathrm{rad}$ was also considered. These simulated results have been compared with the experimental results  \cite{Mazzolari_2024} and it is concluded that the anticlastic radius of the qmBC used in the experiment was about 60 m. 

Simulations were also performed  for different beam divergences (0, 20, 40 and 60 $\mu \mathrm{rad}$). Their comparison with the experimental results \cite{Mazzolari_2024} showed that the best agreement is obtained when the beam divergence is equal to 60 $\mu \mathrm{rad}$. This conclusion confirms the reported experimental observations. Several beam incidence angles were analysed, both in the channeling region (0, -75 and -150 $\mu \mathrm{rad}$) and in the volume reflection region (-300, -375 and -450 $\mu \mathrm{rad}$). This analysis was performed for a BC and a qmBC with $R_\mathrm{a} = 60\ \mathrm{m}$.

The study for different angles of  beam incidence confirms the fine structure in the channeling (volume capture) band of the angular distribution of the deflected positrons. The origin of the fine structure has been thoroughly investigated and explained.

The simulated distributions of deflected positrons and the corresponding experimental results are in good agreement. The normalized root mean square deviation of the angular distribution of the deflected positrons is about 3$\ \%$ for all investigated system configurations. The remaining discrepancies may be due to uncertainties in the geometry of the crystal used in the experiment, e.g. thickness and anticlastic deformation may vary across the beam width. 

Finally, predictions have been made for the crystal orientations for which experimental data are lacking.

\section{Acknowledgements}

We acknowledge support by the European Commission\linebreak{}
through the N-LIGHT Project within
the H2020-MSCA-\linebreak{}RISE-2019 call (GA 872196)
and the EIC Pathfinder Project TECHNO-CLS
(Project No. 101046458). MMM, GRL, \linebreak{}PEIA and JRS would like to acknowledge the support of the national basic and natural sciences program under project code PN223LH010-069.
We also acknowledge the Frankfurt Center for Scientific
Computing (CSC) for providing computer facilities. We are grateful  to thank Dr. A. Mazzolari for providing us with the experimental results discussed in our article and for bringing these findings to our attention.

\section{Authors contributions}
All the authors were involved in the preparation of the
man\-u\-script and contributed equally to this work.
All the authors have read and approved the final manuscript.

\bibliographystyle{unsrturl}
\bibliography{qmBC}

\end{document}